\documentclass{llncs}
\usepackage{graphicx}

\usepackage{makeidx} % allows for indexgeneration
\usepackage[yyyymmdd,hhmmss]{datetime}
\usepackage{color}
\usepackage{amsmath}
\usepackage{amsfonts}
\usepackage{tikz}
\usepackage{algorithm}
\usepackage{algorithmic}
\usepackage{float}
\usepackage{array}
\usepackage{tabulary}
\usepackage{booktabs}
\usepackage{multirow}
\usepackage{textcomp}
\usepackage[title]{appendix}
\usepackage{threeparttable}
\usepackage{amssymb}
\usepackage{mathtools}
\usepackage{url}
\usepackage{kbordermatrix}
\usepackage{mathrsfs} % 2024/04/09 工藤追加
\usepackage{hyperref} % 2024/04/09 工藤追加
\usepackage{stmaryrd} % 2024/04/09 工藤追加
\renewcommand{\kbldelim}{(}
\renewcommand{\kbrdelim}{)}

\usetikzlibrary{intersections, calc, arrows.meta}

\newcommand{\udots}{\mathinner{\mskip1mu\raise1pt\vbox{\kern7pt\hbox{.}} 
 \mskip2mu\raise4pt\hbox{.}\mskip2mu\raise7pt\hbox{.}\mskip1mu}}

\newtheorem{alg}{Algorithm}

\newtheorem{assump}{Assumption}
\newtheorem{heur}{Heuristic}

\renewcommand{\kbldelim}{(}
\renewcommand{\kbrdelim}{)}

\numberwithin{equation}{section}

\def\qed{\hfill $\Box$}

\def\deg{{\rm deg}}

\def\dim{{\rm dim}}
\def\rank{{\rm rank}}

\begin{document}

\title{Polynomial XL: A Variant of the XL Algorithm Using Macaulay Matrices over Polynomial Rings}

\author{Hiroki Furue\inst{1} \and
Momonari Kudo \inst{2}
}

\authorrunning{H. Furue and M. Kudo}

\institute{NTT Social Informatics Laboratories, Tokyo, Japan, \email{hiroki.furue@ntt.com} 
\footnote{This research was conducted while at the University of Tokyo.}
\and
Fukuoka Institute of Technology, Fukuoka, Japan, \email{m-kudo@fit.ac.jp} 
}

% \author{Hiroki Furue}
% \affiliation{%
%   \institution{The University of Tokyo}
%   \city{Tokyo}
%   \country{Japan}
% }
% \email{furue-hiroki261@g.ecc.u-tokyo.ac.jp}

% \author{Momonari Kudo}
% \affiliation{%
%   \institution{The University of Tokyo}
%   \city{Tokyo}
%   \country{Japan}
% }
% \email{kudo@mist.i.u-tokyo.ac.jp}

% \author{
% 	Hiroki Furue \and
% 	Momonari Kudo
% }

% \authorrunning{H. Furue and M. Kudo}

% \institute{
% 	Department of Mathematical Informatics, \\
% 	Graduate School of Information Science and Technology, \\
% 	The University of Tokyo,
% 	Tokyo, Japan \\
% 	\email{furue-hiroki261@g.ecc.u-tokyo.ac.jp \\
% 	kudo@mist.i.u-tokyo.ac.jp}
% }

\maketitle

\begin{abstract}
	
    Solving a system of $m$ multivariate quadratic equations in $n$ variables over finite fields (the MQ problem) is one of the important problems in the theory of computer science.
    The XL algorithm (XL for short) is a major approach for solving the MQ problem with linearization over a coefficient field.
    Furthermore, the hybrid approach with XL (h-XL) is a variant of XL guessing some variables beforehand.
    In this paper, we present a variant of h-XL, which we call the \textit{polynomial XL (PXL)}.
    In PXL, the whole $n$ variables are divided into $k$ variables to be fixed and the remaining $n-k$ variables as ``main variables'', and we generate a Macaulay matrix with respect to the $n-k$ main variables over a polynomial ring of the $k$ (sub-)variables.
    By eliminating some columns of the Macaulay matrix over the polynomial ring before guessing $k$ variables, the amount of operations required for each guessed value can be reduced compared with h-XL.
    Our complexity analysis of PXL (under some \textcolor{black}{practical assumptions and heuristics}) gives a new theoretical bound,
    and it indicates that PXL \textcolor{black}{could be} more efficient than other algorithms in theory on the random system with $n=m$, which is the case {of general multivariate signatures}.
    For example, on systems over the finite field with ${2^8}$ elements with $n=m=80$, the numbers of operations deduced from the theoretical bounds of the hybrid approaches with XL and Wiedemann XL, Crossbred, and PXL with optimal $k$ are estimated as $2^{252}$, $2^{234}$, $2^{237}$, and $2^{220}$, respectively.

	\keywords{MQ problem, MPKC, XL, hybrid approach, Macaulay matrices}

\end{abstract}
%
% This command processes the author and affiliation and title
% information and builds the first part of the formatted document.

%==============================
\section{Introduction}\label{sec:intro}
%==============================

In the field of computer science, the problem of solving a multivariate polynomial system of degree $\geq 2$ over a finite field ({\it the MP problem}) is one of the most important problems, where ``solve'' means to find (at least) one root of the system.
The particular case where polynomials are all quadratic is called {\it the MQ problem}, 
and both the MP and MQ problems
%, where coefficients of polynomials are chosen uniformly and independently at random, 
are known to be NP-hard~\cite{GJ79}.
Moreover, the hardness of the MQ problem is nowadays applied to constructing various cryptosystems (e.g., multivariate public key cryptosystems (MPKCs) such as UOV~\cite{KPG99}).
Therefore, the analysis even for the quadratic case is a very important task both in theory and in practice, and thus we mainly focus on solving the MQ problem in this paper.

%solving the MQ problem is a very important task both in theory and practice, and we mainly focus on solving the MQ problem in this paper.

%In this paper, we mainly focus on solving the MQ problem.
%and both the MP and MQ problems, where coefficients of polynomials are chosen uniformly and independently at random, are known to be NP-hard~\cite{GJ79},
% One of the major applications of the MP and MQ problems is to costruct {\it multivariate public key cryptosystems} (MPKCs), whose security is based on the NP-hardness of these problems.
% MPKCs are expected to be secure against quantum computer attacks, and most of them use only of quadratic polynomials.
% Indeed, the multivariate signature schemes Rainbow~\cite{Rainbow3} and GeMSS~\cite{GeMSS} have been selected in the third round of the NIST post-quantum cryptography standardization project~\cite{NISTpqc}, and their security is based on the difficulty of the MQ problem.
% For this reason, the analysis for solving the MQ problem is a very important task in practice, and we mainly focus on solving the MQ problem in this paper.

A precise definition of the MQ problem is the following:
Let $n$ and $m$ be positive integers, and let $q$ be a power of a rational prime $p$. 
Given a \textcolor{black}{sequence} $F=(f_1, \ldots , f_m)$ of $m$ quadratic polynomials $f_1, \ldots , f_m$ in $n$ variables $x_1, \ldots , x_n$ over a finite field $\mathbb{F}_q$ \textcolor{black}{of $q$ elements}, the MQ problem requires to find \textcolor{black}{at least one} $(a_1, \ldots , a_n)\in \mathbb{F}_q^n$ such that $f_i(a_1, \ldots , a_n) = 0$ for all \textcolor{black}{$i$ with} $1 \leq i \leq m$.
Throughout the rest of this paper, we deal with only the case of $n \le m$ (\textit{overdetermined} case).
This is because algorithms solving the overdetermined MQ problem can be easily applied to the case of $n>m$, since, after \textcolor{black}{the values of} $n-m$ variables are randomly specified, the resulting system will have a solution \textcolor{black}{in most cases}.
Furthermore, this paper evaluates the efficiency of algorithms solving the MQ problem
by substituting specific parameters into the \textcolor{black}{asymptotic} complexity formula
following the security evaluation for various multivariate signatures~\cite{MAYO,UOV,QR-UOV}.

In the literature, there are various methods for solving the MQ problem such as Gr\"{o}bner basis method, Linearization, resultant-based method~\cite[Chapter~3]{CLO_UAG}, and Wu's method~\cite{Wu86}.
In particular, Gr\"obner basis method is a generic method to solve the MQ problem.
The most classical method to compute Gr\"obner bases
is Buchberger's algorithm~\cite{Buch65}, and ones of the currently most efficient algorithms are Faug\`ere's $F_4$ and $F_5$ algorithms~\cite{F99,F02}.
%In 1999, Faug\`ere proposed an efficient variant of Buchberger's algorithm which is called $F_4$ algorithm~\cite{F99},
%and he also published the even more efficient $F_5$ algorithm in 2002~\cite{F02}.$
%The $F_4$ and $F_5$ are currently major algorithms for obtaining a Gr\"obner basis, and solutions of a system can be easily derived from a Gr\"obner basis (\cite[Chapter~3]{CLO_IVA}).
\textcolor{black}{When the ideal generated by $F$ is zero-dimensional, namely the number of (affine) roots of $F$ over an algebraic closure of $\mathbb{F}_q$ is finite}, once a Gr\"{o}bner basis for the input $F$ is computed for a given monomial order (typically a graded reverse lexicographic order is chosen for practical efficiency) with the above algorithms, the FGLM conversion~\cite{FGLM} enables us to obtain its lexicographical Gr\"obner basis, from which \textcolor{black}{roots of $F$} can be easily derived~\cite[Chapter~3]{CLO_IVA}.

As a linearization-based algorithm, Courtois et al.~\cite{CKP00} proposed the \textit{XL algorithm} at EUROCRYPT 2000, and this algorithm is an extension of Relinearization algorithm~\cite{KS99}.
The main idea of XL,
which is already used in~\cite{Laz79,Laz83} by Lazard in order to analyze Buchberger's algorithm, is:
%Lazard used a very similar algorithm alrady in~\cite{} in order to analyze Buchberger's algorithm.
%
%Another major strategy to solve the MQ problem is linearization, and the most basic linearization-based algorithm is the \textit{XL algorithm} proposed by Courtois et al.~\cite{CKP00}.
%The XL algorithm is an extension of Relinearization algorithm~\cite{KS99}, and its idea is:
Linearize the given system by regarding each monomial as one variable, and then, similarly to $F_4$, use linear algebra to the coefficient matrix of the linearized system.
More concretely, we first construct a shift \textcolor{black}{$\mathscr{S}$} of $F$, that is, the set of polynomials of the form $t \cdot f_i$ for all $1 \leq i \leq m$ with monomials $t$ up to given degree.
By linearizing the system \textcolor{black}{defined by $\mathscr{S}$}, we then generate its coefficient matrix (this matrix is nothing but a \textit{Macaulay matrix} of \textcolor{black}{$\mathscr{S}$}), and compute its reduced row echelon form \textcolor{black}{(RREF)} by the row reduction (Gaussian elimination).
If the shift \textcolor{black}{$\mathscr{S}$} is sufficiently large, then the number of \textcolor{black}{linearly} independent polynomials in \textcolor{black}{$\mathscr{S}$} becomes close to the total number of monomials \textcolor{black}{of degree up to the maximal degree of polynomials in $\mathscr{S}$}, and hence a univariate equation would be obtained from \textcolor{black}{RREF} of the Macaulay matrix.
We then solve the obtained univariate equation and repeat such processes with respect to the remaining variables.
Note that XL is considered to be a redundant variant of $F_4$ algorithm (see~\cite{ACF12,AFI04} for details). 
Furthermore, Yang et al.~\cite{YCB07} analyzed a variant of the XL algorithm called \textit{Wiedemann XL (WXL)}, which adopts Wiedemann's algorithm~\cite{W86} instead of row reduction algorithms in the XL framework.
%WXL can be seen as one of the most efficient algorithms solving the $\mathcal MQ$ problem.
WXL provides another complexity estimate \textcolor{black}{that} is used to evaluate the security of various MPKCs such as UOV~\cite{UOV}.

One of the most effective improvements of XL is to apply the {\it hybrid approach}~\cite{BFP09,YCC04} (first proposed as FXL in \cite{YCC04} for XL, in which the ``F'' stands for ``fix''), which is proposed as an approach applying an MQ solver such as $F_4$, $F_5$, or XL efficiently.
This approach fixes the values of $k$ among $n$ variables (say $x_1, \ldots , x_k$), and then solves the remaining system in the $n-k$ variables $x_{k+1}, \ldots , x_n$ using an MQ solver.
These processes are iterated until a solution is found.
%It is known that the $\mathcal MQ$ problem is the hardest in the case of $n=m$.
In the case of $n \approx m$, the hybrid approach may be effective, since the gain obtained by working on systems with less variables may overcome the loss due to the exhaustive search on the fixed variables.
In this paper, we call the hybrid approach with XL (resp.\ WXL) \textit{h-XL} (resp.\ \textit{h-WXL}).
Furthermore, Joux and Vitse proposed the Crossbred algorithm as a practical efficient algorithm for solving MQ systems over the binary field in 2017~\cite{JV17}.
This Crossbred is constructed based on h-XL by eliminating \textcolor{black}{parts} of Macaulay matrices before fixing the \textcolor{black}{values} of some variables. 
In this \textcolor{black}{paper}, we propose a new variant of XL following this direction to further reduce the time complexity.

%============================
\subsubsection*{Our contributions}
%============================
In this paper, we propose a new variant of the XL algorithm, which we call \textit{polynomial XL} (PXL), as an improvement of h-XL.
With notation same as in h-XL described above, the main idea of our improvement is the following:
Before fixing the values of the variables $x_1, \ldots , x_{k}$, we partly perform Gaussian elimination on a Macaulay matrix {\it over the polynomial ring} $\mathbb{F}_q[x_1, \ldots , x_{k}]$, with keeping $x_{1}, \ldots , x_{k}$ as indeterminates.
More specifically, for a given MQ system, \textcolor{black}{namely a sequence $F = (f_1, \ldots , f_m) \in {\mathbb F}_q[x_1, \ldots, x_n]^m$ of $m$ quadratic (not necessarily homogeneous) polynomials $f_1,\ldots, f_m$}, we first regard each $f_i$ as a polynomial in $(\mathbb{F}_q[x_1, \ldots , x_k]) [x_{k+1}, \ldots , x_n]$, and construct a shift of $F$ by multiplying all $f_i$'s by monomials in $x_{k+1}, \ldots , x_n$ (up to some degree).
We then generate the Macaulay matrix $\mathcal{PM}$ of the shift with respect to a {\it graded} monomial order in $x_{k+1}, \ldots , x_n$, where $\mathcal{PM}$ is a {\it polynomial} matrix with entries in the polynomial ring $\mathbb{F}_q[x_1, \ldots , x_k]$.
Here, due to the gradedness of the monomial order, $\mathcal{PM}$ is {\it almost upper-block triangular}, and all of its (nearly-)diagonal blocks are matrices with entries in $\mathbb{F}_q$, {\it not} in $\mathbb{F}_q[x_1, \ldots , x_k]$. 
Thus we can execute row operations on these blocks efficiently, and as a result, we also obtain a partly-reduced matrix.
Under some \textcolor{black}{practical assumption and heuristic} \textcolor{black}{(Assumption \ref{assump:q_semireg} and Heuristic \ref{heur:choose})} such as the semi-regularity of a polynomial sequence, the size of the uneliminated part of this resulting matrix is \textcolor{black}{expected to be} much smaller than that of the original one
\textcolor{black}{(e.g., in the case where $n=m=40$ and $k=10$, the sizes of the original matrix and the uneliminated part are approximately $2^{30}$ and $2^{21}$, respectively)}, \textcolor{black}{so that} the amount of manipulations for each guessed value can be reduced compared with h-XL.
As we will see in Subsection \ref{sub:time_comp} below, this enables us to solve the system
with smaller complexity for some parameters.

We also discuss the time and space complexities \textcolor{black}{of our PXL}, and theoretically compare them with those of h-XL, h-WXL, and Crossbred.
Comparing the time complexities, we show that, {under some practical assumptions and heuristic (\textcolor{black}{Assumptions \ref{assump:semireg_eval} and \ref{assump:q_semireg}}, and Heuristic \ref{heur:choose} below) such as the \textcolor{black}{affine} semi-regularity of \textcolor{black}{polynomial sequences},} our PXL \textcolor{black}{would be} the most efficient in theory for the case of $n \approx m$, see Table~\ref{tab:comp} for details.
For example, on the system over ${\mathbb F}_{2^8}$ with $n=m=80$, the numbers of operations in ${\mathbb F}_q$ required for the execution of h-XL, h-WXL, Crossbred, and PXL are estimated as $2^{252}$, $2^{234}$, $2^{237}$, and $2^{220}$, respectively.
On the other hand, in terms of the space complexity, \textcolor{black}{PXL} might be not well compared to h-WXL since the sparsity of the Macaulay matrix is not maintained through an execution of \textcolor{black}{PXL}.
Therefore, the relationship between PXL and h-WXL can be seen as a trade-off between time and memory.

%=========================
\subsubsection*{Organizations}
%=========================
The rest of this paper is organized as follows:
Section~2 reviews the XL algorithm and the hybrid approach.
Section~3 is devoted to describing the proposed algorithm PXL.
We estimate the time complexity, and theoretically compare it with those of h-XL, h-WXL, and Crossbred in Section~4,
and Section~5 introduces experimental results obtained by our \textcolor{black}{(unoptimized)} implementation of PXL.
Finally, Section~6 is devoted to the conclusion, where we summarize the key points and suggest possible future works.
\textcolor{black}{Also in Appendix \ref{app:semireg}, we recall semi-regular polynomial sequences and their properties.}

%=============================
\section{Preliminaries}\label{sec:pre}
%=============================
In this section, we recall the definition of the XL algorithm~\cite{CKP00}, and discuss its complexity.
We also explain the hybrid approach, which combines an exhaustive search with an MQ solver such as XL.

%=================================
\subsection{Notation and Macaulay matrices}\label{sub:notation}
%=================================

We first fix the notations that are used \textcolor{black}{throughout} the rest of \textcolor{black}{this paper}.
Let $X =\{ x_1,\ldots, x_n \}$ be a set of $n$ variables, and $\mathscr{T}(X)$ denote the set of monomials in $x_1,\ldots , x_n$.
For each non-negative integer $d$, we also denote by $\mathscr{T}(X)_d$ (resp.\ $\mathscr{T}(X)_{\leq d}$) the set of all monomials in $x_1,\ldots,x_n$ of degree $d$ (resp.\ \textcolor{black}{less than or equal to} $d$).
Namely, we set
\begin{align*}
    \mathscr{T}(X) & := \{ x_1^{\alpha_1} \cdots x_n^{\alpha_n} \mid (\alpha_1, \ldots , \alpha_n ) \in (\mathbb{Z}_{\geq 0})^n \},\\
    \mathscr{T}(X)_{d} & := \{ x_1^{\alpha_1} \cdots x_n^{\alpha_n} \in \mathscr{T}(X) \mid \alpha_1 + \cdots + \alpha_n = d\}, \\
    \mathscr{T}(X)_{\leq d} &:= \mathscr{T}(X)_0 \cup \cdots \cup \mathscr{T}(X)_d =\! \{ x_1^{\alpha_1} \!\cdots x_n^{\alpha_n} \in \mathscr{T}(X) \mid \alpha_1 + \cdots + \alpha_n \leq d\} .
\end{align*}
Once $X=\{x_1,\ldots,x_n\}$ is fixed, we may write $\mathscr{T}(X)$, $\mathscr{T}(X)_d$, and $\mathscr{T}(X)_{\leq d}$ as $\mathscr{T}$, $\mathscr{T}_d$, and $\mathscr{T}_{\leq d}$, respectively.
For a commutative ring $A$ of unity, we denote by $A[X] = A[x_1, \ldots , x_n ]$ the polynomial ring with $n$ variables $X =\{ x_1,\ldots,x_n\}$ over $A$.
\textcolor{black}{The total degree of $f \in A[X]$ is denoted by $\deg(f)$, and for a monomial $t \in \mathscr{T}(X)$, let $\mathrm{coeff}(f,t)$ denote the coefficient of $t$ in $f$.}
% Let $R[x] = R[x_1, \ldots , x_n ]$ denote the polynomial ring with $n$ variables over a commutative ring $R$ with unity, and let $\mathrm{Mon}(R[x])$ denote the set of monomials in $R[x]$, say $\mathrm{Mon}(R[x]) = \{ x_1^{\alpha_1} \cdots x_n^{\alpha_n} : (\alpha_1, \ldots , \alpha_n ) \in (\mathbb{Z}_{\geq 0})^n \}$.
% For each $d \geq 0$, we also denote by $\mathscr{T}_d$ (resp.\ $T_{\leq d}$) the set of all monomials in $R[x]$ of degree $d$ (resp.\ $\leq d$).
\textcolor{black}{When $F$ is a set or sequence of polynomials in $A[X]$}, the ideal of \textcolor{black}{$A[X]$} generated by $F$ is denoted by \textcolor{black}{$\langle F \rangle_{A[X]}$} or simply $\langle F \rangle$.
In particular, when $F$ is a finite set $\{ f_1, \ldots , f_m \}$, we denote it by \textcolor{black}{$\langle f_1, \ldots , f_m \rangle_{A[X]}$ or} $\langle f_1, \ldots , f_m \rangle$.
%, namely we set
%\begin{eqnarray}
%	T_d  & = & \left\{ x_1^{\alpha_1} \cdots x_n^{\alpha_n} : \sum_{k=1}^n \alpha_k =d \right\}, \label{eq:mono_d} \\
%	T_{\leq d} &=& T_0 \cup T_{1}\cup \cdots \cup T_d. \label{eq:mono_le_d}
%\end{eqnarray}
For \textcolor{black}{a subset or sequence $F$ of polynomials in $A[X]$, and for a subset $T \subset \mathscr{T}(X)$}, we set $T \cdot F = \{ t \cdot f : t \in T,\ f \in F \}$, which is called the {\it shift} of $F$ by $T$ \textcolor{black}{(we also call a union of shifts a shift).}
\textcolor{black}{As a particular but important case, we define the following shifts:
\begin{align*}
\mathscr{S}_d(F) & := \bigcup_{f\in F_{\leq d}} \mathscr{T}(X)_{d-\deg(f)}\cdot \{f \} =  \{  t f : f \in F_{\leq d},\ t \in \mathscr{T}(X)_{d-\deg(f)}  \}, \\
\mathscr{S}_{\leq d}(F) & := \mathscr{S}_0(F) \cup \cdots \cup \mathscr{S}(X)_d =  \{  t f : f \in F_{\leq d},\ t \in \mathscr{T}(X)_{\leq d-\deg(f)}  \}
\end{align*}
with $F_{\leq d}:= \{ f \in F : \deg(f) \leq d\}$ for each non-negative integer $d$, where ``$\mathscr{S}$'' stands for ``shift''.
In the case where $F_{\leq d}$ is empty, we set $\mathscr{S}_d(F):= \{ 0 \}$ and $\mathscr{S}_{\leq d}(F):= \{ 0 \}$.
We may write $\mathscr{S}_d(F)$ and $\mathscr{S}_{\leq d}(F)$ simply by $\mathscr{S}_d$ and $\mathscr{S}_{\leq d}$ respectively, when $F$ is fixed.}

%=====================================
%\subsection{Macaulay matrices}\label{sub:Mac}
%=====================================

%This subsection reviews the notion of Macaulay matrices, which will be used in the linearization process of the XL algorithm described in Subsection \ref{sub:xl} below.
%For a given system $F=(f_1, \ldots , f_m) \in R[x]^m$, one can linearize it by regarding each monomial as one variable, and then compute the reduced row echelon form of its coefficient matrix (this matrix is nothing but a \textit{Macaulay matrix}).
%its Macaulay matrix is defined as the coefficient matrix for the linearized system.
%Each multivariate polynomial of the system is regarded as a row vector through this linearization, and when $R$ is a field, the $R$-linear space generated by the rows of the Macaulay matrix coincides with that generated by $F$, say $\langle f_1, \ldots , f_m \rangle_R$.

%A precise definition of a Macaulay matrix is as follows:
Here, we recall the definition of Macaulay matrices.
Let \textcolor{black}{$\prec$} be a monomial order on \textcolor{black}{$\mathscr{T}(X)$.
For a sequence $F = (f_1, \ldots, f_m ) \in A[X]^m$ and an ordered subset $T = \{ t_1, \ldots , t_{\ell} \} \subset \mathscr{T}(X)$ with $t_1 \succ \cdots \succ t_{\ell}$,}
%Writing above $F$ and $T$ as $F = \{ f_1, \ldots , f_m \}$ and $T = \{ t_1, \ldots , t_{\ell} \}$ with $t_1 \succ \cdots \succ t_{\ell}$, 
we define the {\it Macaulay matrix} \textcolor{black}{$\mathcal{M}_{\prec}(F,T)$} of $F$ with respect to $T$ as an $(m \times \ell)$-matrix over $R$ whose $(i,j)$-entry is the coefficient of $t_j$ in $f_i$, say
\begin{eqnarray*}
  {\color{black}\mathcal{M}_{\prec}}(F,T) :=
  \kbordermatrix{
        &t_1& \cdots & t_{\ell}\\
     f_1& \mathrm{coeff}(f_1,t_1) & \cdots & \mathrm{coeff}(f_1,t_{\ell}) \\
    \vdots & \vdots & & \vdots \\
    f_m& \mathrm{coeff}(f_m,t_1) & \cdots & \mathrm{coeff}(f_m,t_{\ell}) \\
    }.
\end{eqnarray*}
When \textcolor{black}{$\prec$} is clear from the context, we simply denote it by \textcolor{black}{$\mathcal{M}(F,T)$}.

Conversely, for an $(m \times \ell)$-matrix \textcolor{black}{$M=(a_{i,j})$} over \textcolor{black}{$A$} and for $T$ given as above, let \textcolor{black}{$\mathcal{M}_{\prec}^{-1}(M,T)$ (or $\mathcal{M}^{-1}(M,T)$ simply)} denote a unique list $F'$ of polynomials in \textcolor{black}{$A[X]$} such that \textcolor{black}{$\mathcal{M}_{\prec} (F', T) = M$}, namely, we set $g_i := \sum_{j=1}^{\ell} a_{i,j} t_j$ for $1 \leq i \leq m$, and \textcolor{black}{$\mathcal{M}_{\prec}^{-1}(M,T) := ( g_1, \ldots , g_{m} )$}.

\begin{example}
	Consider the following three quadratic polynomials (over $R=\mathbb{Z}$) in two variables $x_1$ and $x_2$:
	\begin{eqnarray*}
		f_1 &=& 5 x_1^2 + 6 x_1 x_2 + 4 x_1 + 5 x_2 + 3, \\
		f_2 &=& 4 x_1^2 + 5 x_1 x_2 + 3 x_2^2 + 6 x_1 + 2 x_2 + 2, \\
		f_3 &=& 2 x_1^2 + 4 x_1 x_2 + 2 x_2^2 + 6 x_1 + x_2 + 2.
	\end{eqnarray*}
	When we put $F := ( f_1, f_2, f_3 )$, we construct a Macaulay matrix of the shift \textcolor{black}{$\mathscr{S}_{3} = \mathscr{S}_3(F) = \mathscr{T}_{1} \cdot F = \{ x_i f_j : 1 \leq i \leq 2,\, 1 \leq j \leq 3 \}$}, where \textcolor{black}{$\mathscr{T}_{1}$} is the set of monomials in $x_1$ and $x_2$ of degree one.
	We order elements of \textcolor{black}{$\mathscr{S}_3$} as follows:
	${\color{black}\mathscr{S}_3} = \{ x_1 f_1, x_1 f_2, x_1 f_3, x_2 f_1, x_2 f_2, x_2 f_3 \} $.
	Let \textcolor{black}{$\prec_{\rm glex}$} be the graded \textcolor{black}{lexicographic} order on the monomials in $x_1$ and $x_2$ with $x_1 \succ x_2$, that is, $x_1^{\alpha_1} x_2^{\alpha_2} {\color{black}\prec_{\rm glex}} x_1^{\beta_1} x_2^{\beta_2}$ if \textcolor{black}{$\alpha_1 + \alpha_2 < \beta_1 + \beta_2$}, or \textcolor{black}{$\alpha_1 + \alpha_2 = \beta_1 + \beta_2$ and $x_1^{\beta_1} x_2^{\beta_2}$ is greater than $x_1^{\alpha_1} x_2^{\alpha_2}$} with respect to the lexicographical order with $x_1 \succ x_2$.
	When we order elements of \textcolor{black}{$\mathscr{T}_{\leq 3}$} (which is the set of monomials in \textcolor{black}{$X = \{ x_1,x_2\}$} of degree $\leq 3$) by \textcolor{black}{$\prec_{\rm glex}$}, the Macaulay matrix \textcolor{black}{$\mathcal{M}_{\prec_{\rm glex}}(\mathscr{S}_3,\mathscr{T}_{\leq 3})$} of \textcolor{black}{$\mathscr{S}_3$} with respect to \textcolor{black}{$\mathscr{T}_{\leq 3}$} is given as follows:
	\begin{eqnarray*}
	\mathcal{M}_{\prec_{\rm glex}}(\mathscr{S}_{3},\mathscr{T}_{\leq 3}) =
			\kbordermatrix{
				& x_1^3 & x_1^2 x_2 & x_1 x_2^2 & x_2^3 & {x_1^2} & {x_1 x_2} & {x_2^2} & {x_1} & {x_2} & {1} \\
				{x_1f_1}&5&6&0&0&4&5&0&3&0&0 \\
				{x_1f_2}&4&5&3&0&6&2&0&2&0&0 \\
				{x_1f_3}&2&4&2&0&6&1&0&2&0&0 \\
				{x_2f_1}&0&5&6&0&0&4&5&0&3&0 \\
				{x_2f_2}&0&4&5&3&0&6&2&0&2&0 \\
				{x_2f_3}&0&2&4&2&0&6&1&0&2&0 \\
                %
    %             {f_1}&0&0&0&0&5&6&0&4&5&3 \\
				% {f_2}&0&0&0&0&4&5&3&6&2&2 \\
				% {f_3}&0&0&0&0&2&4&2&6&1&2 \\
			}.
	\end{eqnarray*}
\end{example}

In the XL algorithm in Subsection \ref{sub:xl}, the reduced row echelon form of a Macaulay matrix of a shift of $F$ is computed, with $R$ a finite field $\mathbb{F}_q$ \textcolor{black}{of order $q$, where $q$ is a power of a prime}.
This corresponds to computing a basis $G$ of the $\mathbb{F}_q$-vector space generated by the shift, and clearly the computed basis also generates the ideal \textcolor{black}{$\langle F \rangle_{\mathbb{F}_q[X]}$, i.e., $\langle G \rangle_{\mathbb{F}_q[X]} = \langle F \rangle_{\mathbb{F}_q[X]}$}.
In general, $G$ computed as above is not necessarily a Gr\"{o}bner basis of \textcolor{black}{$\langle F \rangle_{\mathbb{F}_q[X]}$}, but we will review in \textcolor{black}{Subsection \ref{subsec:DB} below} that for sufficiently large shifts, $G$ becomes a Gr\"{o}bner basis.

%=======================================
\subsection{XL algorithm}\label{sub:xl}
%=======================================

This subsection briefly reviews {\it the XL algorithm} (which stands for eXtended Linearizations), which is proposed in~\cite{CKP00} by Courtois et al.\ to find a solution to a system of multivariate polynomials over finite fields.
We write down the XL algorithm in Algorithm \ref{alg:XL} below, where the notations are the same as in the previous subsections.
We also suppose that the input system is zero-dimensional, \textcolor{black}{namely, the input system has only finite (affine) roots over an algebraically closed field}.
Note also that the input polynomials are assumed to be all quadratic as in the original paper~\cite{CKP00}, but in fact, their idea is applicable to a general multivariate system of higher degree.

%\newpage
\begin{alg}[XL, {\cite[Section 3, Definition 1]{CKP00}}]\hspace{-5pt}{\bf{.}}\label{alg:XL} 
~
\begin{description}
\item[{\it Input:}] \textcolor{black}{A sequence} $F= ( f_1, \ldots , f_m ) \in \mathbb{F}_q[x_1, \ldots, x_n]^m$ \textcolor{black}{of (not necessarily homogeneous) quadratic polynomials}, and \textcolor{black}{a natural number} $D$ \textcolor{black}{with $D \geq 2$}.
\item[{\it Output:}] A solution over $\mathbb{F}_q$ to $f_i(x_1, \ldots , x_n) = 0$ for $1 \leq i \leq m$.
\end{description}
\begin{enumerate}
\item \textbf{Multiply:} Computing all the products $t \cdot f_i$ with $t \in {\color{black}\mathscr{T}_{\leq D-2}}$, construct the shift \textcolor{black}{$\mathscr{S}_{\le D} := \mathscr{S}_{\leq D}(F) = \mathscr{T}_{\leq D-2} \cdot F$, which is the shift of $F$ by $\mathscr{T}_{\leq D-2}$}.
\item \textbf{Linearize:} Make the Macaulay matrix \textcolor{black}{$M := \mathcal{M}_{\prec}(\mathscr{S}_{\leq D}, \mathscr{T}_{\leq D})$} with respect to some elimination monomial order \textcolor{black}{$\prec$} such that all the terms containing one variable (say $x_n$) are eliminated last.
Compute the reduced row echelon form $B$ of \textcolor{black}{$M$}, and put \textcolor{black}{$G:= \mathcal{M}_{\prec}^{-1} (B, \mathscr{T}_{\leq D})$}.
A univariate polynomial $g(x_n)$ in $x_n$ of degree at most $D$ is surely contained in $G$ when $D$ is sufficiently large.
%(see Subsection \ref{subsec:DB} or Appendix \ref{sec:correctness} for details).
\item \textbf{Solve:} Compute the roots in $\mathbb{F}_q$ of $g$ by e.g., combining square-free, distinct-degree and equal-degree factorization algorithms such as {\rm \cite{Yun76}}, {\rm \cite{KS98}} and {\rm \cite{GS92}} respectively.
\item \textbf{Repeat:} Substitute a root into $x_n$, simplify the equations of $G$, and then find the values of the other variables.
\end{enumerate}
\end{alg}
Note that in the generation of \textcolor{black}{$\mathcal{M}_{\prec}(\mathscr{S}_{\leq D}, \mathscr{T}_{\leq D})$}, one can sort elements in $\mathscr{S}_{\leq D}$ arbitrarily.
\textcolor{black}{We also note that, in XL, it suffices to obtain a univariate polynomial in Step (3) to continue the procedures, whence we do not need to compute a Gr\"{o}bner basis.}
\textcolor{black}{On the other hand, XL can be described as a redundant variant of $F_4$, supposing an assumption that the input system $F$ has only one solution over a finite field, see \cite{AFI04} for details.
Moreover, we remark that we can use any other monomial order (e.g., a graded monomial order), if we execute only Steps (1) and (2) to obtain a Gr\"{o}bner basis of $\langle F \rangle$ (in this case, the computation can be viewed as a special case of Lazard's algorithm~\cite{Laz79,Laz83}).
Even in this case, we can obtain a root easily from the computed Gr\"{o}bner basis, under an assumption similar to \cite{AFI04}, see Remark \ref{rem:gradedorder} below for details.}

The condition of the \textcolor{black}{natural number $D$} for \textcolor{black}{XL to continue the procedures} is discussed in the next subsection.

%==============================================
\subsection{Degree bounds for the success of XL}\label{subsec:DB}
%==============================================

Algorithm \ref{alg:XL} has an input parameter $D$ called a \textcolor{black}{\it degree bound}, and it is known that the algorithm surely finds a zero of $\langle F \rangle$ for sufficiently large $D$.
This subsection reviews bounds on such $D$ both in theory and in practice.
Let $R := K[x_1,\ldots,x_n]$ be the polynomial ring of $n$ variables over a field $K$, and $F = (f_1,\ldots, f_m)$ be a sequence of {\it not necessarily homogeneous} polynomials in $R$ of positive degrees $d_1, \ldots d_m$, respectively.
We denote by $f^{\rm top}$ the maximal homogeneous part of $f \in R \smallsetminus \{ 0 \}$, and put $F^{\rm top}:= (f_1^{\rm top},\ldots, f_m^{\rm top})$.
Put $R' = R[y]$ for an extra variable $y$ for homogenization.
We also denote by $f^h$ the homogenization of $f \in R\smallsetminus \{ 0 \}$ by $y$, say $f^h = y^{\deg(f)}f(x_1/y,\ldots, x_n/y)$, and put $F^h := (f_1^h,\ldots, f_m^h) \in (R')^m$.
\textcolor{black}{For each $d \in \mathbb{Z}$, let $I_d$ denote the degree-$d$ homogeneous component of a homogeneous ideal $I$ of $R$ (resp.\ $R'$), namely $I_d = I \cap R_d$ (resp.\ $I_d = I \cap (R')_d$).
We put $I_{\leq d} := I \cap R_{\leq d}$ with $R_{\leq d} := \bigoplus_{i=0}^d R_i$ for a (not necessarily homogeneous) ideal $I$ of $R$, and this kind of notation is applied to $R'$ and its arbitrary ideal.}

% It will be proved in Corollary \ref{cor:main} (in Appendix \ref{sec:correctness}) that Algorithm \ref{alg:XL} in Step (2) outputs a Gr\"{o}bner basis of $\langle F \rangle$ with respect to an elimination order, for $D$ larger than {\it Dub\'{e}'s degree bound} $2\left( \left( d^2/2 \right) +d \right)^{2^{n-1}}$, where $d := \mathrm{max}\{ \mathrm{deg}(f_i) : 1\leq i \leq m \}$. 
% This implies that for such $D$, Algorithm \ref{alg:XL} surely finds a zero of $\langle F \rangle$, see Proposition~\ref{prop:correctness}.
%
A well-known (theoretical) upper bound is {\it Dub\'{e}'s degree bound}~\cite{Dube}
given by $D(n,d) := 2\left( \left( d^2/2 \right) +d \right)^{2^{n-1}}$ with $d := \mathrm{max}\{ \mathrm{deg}(f_i) : 1\leq i \leq m \}$.
%$\mathrm{Dube}_{n+1,d}$~\cite{Dube}, where $d := \mathrm{max}\{ \mathrm{deg}(f_i) : 1\leq i \leq m \}$. 
\textcolor{black}{For any degree $D$ larger than or equal to the Dub\'{e}'s bound, the} reduced row echelon form of \textcolor{black}{$\mathcal{M}_{\prec}(\mathscr{S}_{\leq D}, \mathscr{T}_{\leq D})$ with $\mathscr{S}_{\leq D}= \mathscr{S}_{\leq D}(F)$ and $\mathscr{T}_{\leq D}=\mathscr{T}(X)_{\leq D}$} yields a Gr\"{o}bner basis of $\langle F \rangle$ with respect to an elimination order \textcolor{black}{$\prec$}.
%see Corollary~\ref{cor:AL} in Appendix \ref{sec:correctness}.
%$D \geq \mathrm{Dube}_{n+1,d}$.
Hence, for such a $D$ one can obtain a \textcolor{black}{root} of $F$ with Algorithm \ref{alg:XL}.
%(cf.\ Proposition \ref{prop:correctness} below).

However, \textcolor{black}{Dub\'{e}'s degree bound would be impractical} \textcolor{black}{under the cryptographic setting,} and we \textcolor{black}{here} recall \textcolor{black}{quite smaller bounds under the following assumption:}

\begin{assump}\label{assump:semireg}\hspace{-5pt}{\bf{.}}
\textcolor{black}{The input sequence $F=(f_1,\ldots,f_m)$ is {\it affine semi-regular}, namely $F^{\rm top} = (f_1^{\rm top},\ldots, f_m^{\rm top})$ is semi-regular.}
\end{assump}

\textcolor{black}{See Definition \ref{def:affine_semireg} in Appendix \ref{app:semireg} below for the definition of affine semi-regular sequences.}
Semi-regular sequences are important in the theory of solving polynomial systems (cf.\ \cite{B04}, \cite{BFSY}), and often (e.g., \cite[Section 4.3]{ikematsu2023recent}) the security of multivariate cryptosystems \textcolor{black}{is evaluated} under Assumption \ref{assump:semireg}.
% This assumption has been considered to be practical in the cryptographic literature} \textcolor{black}{(cf.\ \cite{B04})}.
\textcolor{black}{Under Assumption \ref{assump:semireg}, a bound for the success of XL is obtained by considering the rank of the Macaulay matrix $\mathcal{M}_{\prec}(\mathscr{S}_{\le d},\mathscr{T}_{\le d})$, denoted by $\rank(\mathcal{M}_{\prec}(\mathscr{S}_{\le d},\mathscr{T}_{\le d}))$, where $\mathscr{S}_{\leq d}= \mathscr{S}_{\leq d}(F)$ and $\mathscr{T}_{\leq d}=\mathscr{T}(X)_{\leq d}$ \textcolor{black}{with $X = \{ x_1,\ldots,x_n\}$}.
This rank is clearly equal to the dimension $\mathrm{dim}_{K}(\langle \mathscr{S}_{\le d}(F) \rangle_{K})$ of the $K$-vector space $\langle \mathscr{S}_{\leq d}(F) \rangle_K$ generated by $\mathscr{S}_{\leq d}(F)$, and it does not depend on the order of the monomials in $\mathscr{T}_{\leq d}$.
Thus, we need to investigate $\mathrm{dim}_{K}(\langle \mathscr{S}_{\le d}(F) \rangle_{K})$.
For this, let us first recall the following theorem, whose mathematically rigorous and correct proof is given in \cite{KY} (or \cite{KY2}) by Kudo-Yokoyama:} 
%a practical bound given in \cite{YC04} in the following.
%On the other hand, a practical bound is given in \cite{YC04}, and  it tends to be {\it much} smaller than the Dub\'{e}'s one.

\begin{theorem}[{\cite[Theorem 1 \& 7, Corollary 1]{KY}, \cite[Theorem 1]{KY2}}]\label{th:KY}
    \textcolor{black}{
    % Let $R = K[x_1,\ldots, x_n]$ be the polynomial ring with $n$ variables over a field $K$, and put $R' = R[y]$ for an extra variable $y$ for homogenization.
    % With notation as above, put $R' = R[y]$ for an extra variable $y$ for homogenization.
    % We also denote by $f^h$ the homogenization of a non-zero polynomial $f \in R$ by $y$, say $f^h = y^{\deg(f)}f(x_1/y,\ldots, x_n/y)$.
    % Let $F = (f_1,\ldots, f_m) \in R^m$ be a sequence of polynomials of positive degrees, and 
    With notation as above, assume that the sequence $F=(f_1,\ldots,f_m)$ of not necessarily homogeneous polynomials satisfies Assumption \ref{assump:semireg}.
    Let $d_{\rm reg}(F^{\rm top})$ denote the degree of regularity for the homogeneous ideal $\langle F^{\rm top} \rangle_R$, defined as in Definition \ref{def:dreg}.}
    %is affine semi-regular (i.e., $F^{\rm top} = (f_1^{\rm top},\ldots, f_m^{\rm top})$ is semi-regular).
    Then, for any non-negative integer $d$ with $d< d_{\rm reg}(F^{\rm top})$, we have
     \begin{equation*}
             \dim_K (R')_d/ \langle F^h \rangle_d = \sum_{i=0}^d \dim_K R_d / \langle F^{\rm top}\rangle_d
     \end{equation*}
    with $F^h: = (f_1^h,\ldots, f_m^h)$.
    Hence, the Hilbert series $\mathrm{HS}_{R'/ \langle F^h \rangle}(z)$ of $R'/ \langle F^h \rangle$ satisfies
     \begin{equation*}
         \mathrm{HS}_{R'/ \langle F^h \rangle}(z) \equiv \frac{\prod_{j=1}^m (1-z^{d_j})}{(1-z)^{n+1}} \pmod{z^{D}}
     \end{equation*}
     for $d_j := \deg (f_j)$ and $D:=d_{\rm reg}(F^{\rm top})$, so that $F^h$ is $d_{\rm reg}(F^{\rm top})$-regular.
    Moreover, if $d_{\rm reg}(F^{\rm top})<\infty$ (which is equivalent to $m \geq n$ under Assumption \ref{assump:semireg}), then the number of projective zeros of $\langle F^h \rangle_{R'}$ is finite at most, whence $\langle F \rangle_R$ is zero-dimensional.
\end{theorem}

\textcolor{black}{
Note that $d_{\rm reg}(F^{\rm top})$ in Theorem \ref{th:KY} is easily computed from the Hilbert series given in \eqref{eq:semiregHil2}, and in fact it does not depend on $F^{\rm top}$ but is determined only by $n$, $m$, and $d_1,\ldots, d_m$.
From this, for fixed $m$ and $d_1,\ldots,d_m$, we set
\[
D_{\rm reg}^{(n)} := d_{\rm reg}(F^{\rm top}) = \min \left\{ \; d \; \left| \;  \mathrm{coeff}\left( \frac{\prod_{j=1}^{m}(1-z^{d_j})}{(1- z)^n} ,t^d \right) \le 0 \right. \right\},
\]
which we interpret as $\infty$ if $m < n$.
In particular, if $d_1 = \cdots = d_m = 2$, we have
\begin{equation*}
		\label{eq:regularity_0}
		D_{\rm reg}^{(n)} = \min \left\{ \; d \; \left| \; \mathrm{coeff}\left( \left( 1-z \right)^{m-n} \left( 1+z \right)^m ,t^d \right) \le 0  \right. \right\}.
\end{equation*}
Here, even if we do not suppose the affine semi-regularity of $F$, we have
\[
\langle F^h \rangle_d = \langle \mathscr{S}_d (F^h) \rangle_K \cong \langle \mathscr{S}_{\leq d}(F) \rangle_K \subset \langle F \rangle_{\leq d} 
\]
as $K$-vector spaces, where a $K$-isomorphism is given by the dehomogenization map $\langle \mathscr{S}_d (F^h) \rangle_K \ni h \longmapsto h|_{y=1} \in \langle \mathscr{S}_{\leq d}(F) \rangle_K$ (see e.g., \cite[Section 4]{Diem04} for details), and therefore
\[
    \dim_K \langle F^h \rangle_d = \dim_K \langle \mathscr{S}_{d}(F^h) \rangle_K = \dim_K \langle \mathscr{S}_{\leq d}(F) \rangle_K \leq \dim_K \langle F \rangle_{\leq d}.
\]
Moreover, it follows that $\dim_K (R')_d = |\mathscr{T}(X\cup\{y\})_d| = \dim_K R_{\leq d} = |\mathscr{T}_{\leq d}|$.
Hence, as a corollary of Theorem \ref{th:KY}, we obtain the following:}

\begin{corollary}[cf.\ {\cite[Proposition 1]{YC04}}]\label{cor:deg}
\textcolor{black}{	
 Under the same setting and assumptions as in Theorem \ref{th:KY}, for any $d$ with $d < D_{\rm reg}^{(n)} = d_{\rm reg}(F^{\rm top})$, we have
	\begin{equation*}\label{eq:dim}
		|\mathscr{T}_{\le d}| - \mathrm{dim}_{K}(\langle \mathscr{S}_{\le d}(F) \rangle_{K}) = \mathrm{coeff}\left( \frac{\prod_{j=1}^m (1-z^{d_j})}{(1-z)^{n+1}},z^d \right).
	\end{equation*}
 In particular, if the elements of $F$ are all quadratic, then we have
		\begin{equation*}\label{eq:dim_quad}
		|\mathscr{T}_{\le d}| - \mathrm{dim}_{K}(\langle \mathscr{S}_{\le d}(F) \rangle_{K}) = \mathrm{coeff}\left( \left( 1-z \right)^{m-n-1} \left( 1+z \right)^m ,z^d \right)
	\end{equation*}
 for any $d$ with $d < D_{\rm reg}^{(n)}$.}
\end{corollary}

% \begin{remark}
%     \textcolor{black}{If the }
% \end{remark}

In the context of the above discussion, we here list the following \textcolor{black}{two} kinds of bounds on $D$ for which Algorithm \ref{alg:XL} finds a solution:

\subsubsection*{Heuristic but practical bound from Yang-Chen, \textcolor{black}{Ars et al.}, and Diem's studies.}
\textcolor{black}{Assuming that $F$ is an affine semi-regular sequence of quadratic polynomials, we consider a sufficient condition that a univariate polynomial in $x_n$ is obtained in Step (2) of Algorithm \ref{alg:XL}, when we use an elimination order such that $x_n^D,x_n^{D-1},\dots,x_n,1$ are listed at the end.
It is straightforward that the last non-zero row vector of the reduced row echelon form of $\mathcal{M}(\mathscr{S}_{\le D},\mathscr{T}_{\le D})$ yields a univariate equation of $x_n$ if $\rank(\mathcal{M}(\mathscr{S}_{\le D},\mathscr{T}_{\le D}))$ is larger than the number of columns minus $D+1$, i.e.,
\begin{equation*}\label{eq:D}
\rank(\mathcal{M}(\mathscr{S}_{\le D},\mathscr{T}_{\le D})) \geq |\mathscr{T}_{\le D}|-D,
\end{equation*}
equivalently
\begin{equation}\label{eq:Deq}
\chi(D):= |\mathscr{T}_{\le D}| - \mathrm{dim}_{K}(\langle \mathscr{S}_{\le D}(F) \rangle_{K}) \leq D,
\end{equation}
which is used in \cite{Diem04} and \cite{mcguire2021termination}.
Thus, it follows from Corollary \ref{cor:deg} that the minimum $D$, denoted by $D_{\rm XL}$ here, required for the success of Step (2) of Algorithm \ref{alg:XL} is upper-bounded by}
\begin{equation}
\label{eq:dreg}
	D_{\rm XL} \leq D_0 := \min \left\{ \; d \; \left| \;  \mathrm{coeff}\left( \left( 1-z \right)^{m-n-1} \left( 1+z \right)^m ,z^d \right) \le d \right. \right\}
\end{equation}
\textcolor{black}{if $D_{\rm XL} < D_{\rm reg}^{(n)}$.
The condition $\mathrm{coeff}( ( 1-z)^{m-n-1} ( 1+z )^m ,z^d) \le d$ is equivalent to that the $z^d$-coefficient of $(1-z)^{m-n-1}(1+z)^m - (1-z)^{-2}$ is negative (cf.\ \cite[Section 5.1]{AFI04}).}
Note that, even when $D_{\rm XL} \geq D_{\rm reg}^{(n)}$, it would be possible that Step (2) of Algorithm \ref{alg:XL} produces a univariate polynomial at the degree equal to this upper-bound:
See \cite[Section 4]{YC04}, where the authors of \cite{YC04} say ``the minimum $D$ required for the reliable termination of XL is given by $D_0$''.
From this, we may estimate $D_{\rm XL} \approx D_0$.
\textcolor{black}{Assuming the {\it Maximum Rank Conjecture} (which is equivalent over an infinite field to Fr\"{o}berg conjecture~\cite{Fro}, see \cite{Pardue} for a proof of the equivalency),} \textcolor{black}{Diem also proved in \cite[Theorem 1]{Diem04} that $D_0$ is a lower bound for \eqref{eq:Deq} to be satisfied.}
%in the case where $q$ is sufficiently large.
% \textcolor{black}{In this paper, we} call this \textcolor{black}{bound $D_0$ (i.e., the right hand side of \eqref{eq:dreg})} the \textit{solving degree of XL} \textcolor{black}{(in quadratic case)}.
% Note that if $n$ and $m$ are determined, then the solving degree can be computed.
One can easily confirm that \textcolor{black}{$D_0$} tends to be much smaller than Dub\'{e}'s degree bound (e.g., \textcolor{black}{the value of $D_0$} on systems with $n=10$ and $m=11$ is $11$, whereas Dub\'{e}'s degree bound on the same system is approximately $10^{309}$).

\begin{remark}\label{rem:gradedorder}
    In the case where we use a graded monomial order as noted in Subsection \ref{sub:xl}, we consider the inequality $\chi (D) \leq 1$ instead of \eqref{eq:Deq} as a sufficient condition for XL to compute a solution, supposing the following (i) and (ii):
    \begin{enumerate}
        \item[(i)] $F$ has at most one root (counted with multiplicity) over an \textcolor{black}{algebraic} closure $\overline{K}$ of $K$ (cf.\ \cite[Condition 1]{AFI04} for a similar condition).
        \item[(ii)] $F^{\rm top}$ has no root other than $(0,\ldots,0)$.
    \end{enumerate}
    Under these assumptions, there exists a sufficiently large integer $d$ such that the above inequality definitely holds for any $D$ with $D \geq d$.
    Indeed, it follows from (i) and (ii) that the number of projective zeros over $\overline{K}$ of $F^h$ is also finite (in fact one at most), whence there exists $d>0$ such that for any $D$ with $D \geq d$, the value of the Hilbert function ${\rm HF}_{R'/\langle F^h \rangle}(D)=\dim_K (R'/\langle F^h \rangle)_D = \chi(D)$ is equal to the number of roots (counted with multiplicity) over $\overline{K}$ of $F$, \textcolor{black}{see e.g., \cite[Proposition 3.3.6]{Capaverde14} or \cite[Corollary 10]{YC04} for a proof (see also \cite[Lemma 2.2.2]{KY2})}.
    In this case, we remark that the reduced Gr\"{o}bner basis of $\langle F \rangle$ is $\{ x_1-a_1,\ldots,x_n-a_n \}$, where $(a_1,\ldots,a_n)$ is the unique root of $F$.
    \textcolor{black}{For $m > n$,} we estimate
    \begin{equation}\label{eq:DXL}
        D_{\rm XL} \approx D_1 := \min \left\{ d \geq 2 \left| \;  \mathrm{coeff}\left( \left( 1-z \right)^{m-n-1} \left( 1+z \right)^m ,t^d \right) \le 1 \right. \right\},
    \end{equation}
     by a discussion following \cite[Section 4]{YC04}, similarly to the case of elimination order.
     \textcolor{black}{Note that the cases $d= 0$ and $d = 1$ are removed in \eqref{eq:DXL}, since $\chi(0)=1$ and $\chi(1) = n+1$.
     We also note that $D_1 \geq D_0$.}
     We experimentally confirmed that, in most cases, XL for $D = D_1$ computes a Gr\"{o}bner basis of the input system:
     In our experiments, we randomly generated sequences $H = (h_1,\ldots,h_m)$ of quadratic non-homogeneous polynomials over $\mathbb{F}_{31}$ with no constant term for several small $n$ and for all $m$ with $n<m \leq 2n$.
     For each generated sequence $H$, we choose $(a_1,\ldots,a_n) \in \mathbb{F}_{31}^n$ at random, and then put $f_i := h_i(x_1,\ldots,x_n) - h_i(a_1,\ldots,a_n)$ for $1\leq i \leq m$ and $F := (f_1,\ldots,f_m)$.
     Then each sequence $F$ constructed as above would satisfy the above assumptions (i) and (ii) (in fact, $F^{\rm top}$ would be semi-regular) with its unique root $(a_1,\ldots,a_n)$, in most cases.
     This construction of $H$ and $F$ may correspond to the general construction of multivariate public key encryption (see e.g., \cite[Section 2.2]{DPS}, \cite[Section 4.3]{ikematsu2023recent}).
     % (i) $F^{\rm top}$ is semi-regular (in particular, $d_{\rm reg}(F^{\rm top}) < \infty$, so that $F^{\rm top}$ has no root other than $(0,\ldots,0)$), (ii) $F$ has only one affine root $(0,\ldots, 0)$, and (iii) the reduced Gr\"{o}bner basis of $F$ is $\{x_1,\ldots,x_n\}$.
     % For an arbitrary point $(a_1',\ldots,a_n') \in K^n$, we could obtain an affine semi-regular sequence $F' = (f_1',\ldots, f_m')$ having the point as a unique root by putting $f_i' := f_i(x_1,\ldots,x_n)-b_i$ with $b_i := f_i(a_1',\ldots,a_n')$.
     %applying linear coordinate exchange to $F$, and the reduced Gr\"{o}bner basis of $F'$ is $\{ x_1 - b_1,\ldots,x_n - b_n\}$ as in \cite[Section 4.2]{AFI04}.
     Therefore, our experiments would be meaningful.
     %From the above discussion, we obtain the following heuristic:}
     % In our experiments, we randomly generated sequences $F = (f_1,\ldots,f_m)$ of quadratic non-homogeneous polynomials over $\mathbb{F}_{31}$ with no constant term for several small $n$ and for all $m$ with $n<m \leq 2n$. 
     % Note that each sequence $F$ constructed as above would satisfy the above assumptions (i) and (ii) (in fact, $F^{\rm top}$ would be semi-regular), where $F$ has a unique root $(a_1,\ldots,a_n)=(0,\ldots,0)$.
     % % (i) $F^{\rm top}$ is semi-regular (in particular, $d_{\rm reg}(F^{\rm top}) < \infty$, so that $F^{\rm top}$ has no root other than $(0,\ldots,0)$), (ii) $F$ has only one affine root $(0,\ldots, 0)$, and (iii) the reduced Gr\"{o}bner basis of $F$ is $\{x_1,\ldots,x_n\}$.
     % For an arbitrary point $(a_1',\ldots,a_n') \in K^n$, we could obtain an affine semi-regular sequence $F' = (f_1',\ldots, f_m')$ having the point as a unique root by putting $f_i' := f_i(x_1,\ldots,x_n)-b_i$ with $b_i := f_i(a_1',\ldots,a_n')$.
     % %applying linear coordinate exchange to $F$, and the reduced Gr\"{o}bner basis of $F'$ is $\{ x_1 - b_1,\ldots,x_n - b_n\}$ as in \cite[Section 4.2]{AFI04}.
     % Therefore, the above construction of $F$ would be meaningful.
     %From the above discussion, we obtain the following heuristic:}
\end{remark}

\subsubsection*{Expected theoretical bound from Semaev-Tenti and Kudo-Yokoyama's results.}
\textcolor{black}{
We also note that, as a theoretical upper-bound on $D_{\rm XL}$, we may apply the following upper-bound on the {\it solving degree} of Gr\"{o}bner basis computation:}

\begin{theorem}[{\cite[Lemma 4]{KY}}, {\cite[Theorem 3]{KY2}}]\label{th:KY2}
    \textcolor{black}{Let $F = (f_1,\ldots, f_m)$ be a (not necessarily semi-regular) sequence of polynomials in $K[x_1,\ldots,x_n]$, and $\prec$ be a graded reverse lexicographic order on the monomials in $x_1,\ldots, x_n$.
    If $d_{\rm reg}(F^{\rm top}) < \infty$, then there constructively exists a Buchberger-like algorithm $\mathcal{A}$ for computing a Gr\"{o}bner basis for $F$ with respect to $\prec$ such that the degree of critical S-pairs (resp.\ S-polynomials) appearing in the execution of $\mathcal{A}$ is upper-bounded by $2 d_{\rm reg}(F^{\rm top}) - 1$ (resp.\ $2 d_{\rm reg}(F^{\rm top}) - 2$).}
\end{theorem}

\textcolor{black}{
These upper-bounds had been proved by Tenti in his PhD thesis~\cite[Theorem 3.65]{Tenti} (see also \cite[Theorem 2.1]{ST} by Semaev-Tenti) under some constraints (e.g., $F$ contains field equations $x_i^q-x_i$ for $1\leq i \leq n$), and Kudo-Yokoyama extended his result to a general case in \cite[Section 5]{KY} (see also \cite[Section 4]{KY2} for algorithmic details).
Since we can interpret the Gr\"{o}bner basis computation as repeating to execute row reductions on Macaulay matrices as in $F_4$~\cite{F99} and (matrix-)$F_5$~\cite{F02}, we {\it expect} that $D_{\rm XL} \leq 2 d_{\rm reg}(F^{\rm top}) - 1$.
As for the magnitude relation between \textcolor{black}{$D_1$} and $2 d_{\rm reg}(F^{\rm top}) - 1$, they are not equal to each other in general, and both \textcolor{black}{$D_1 < 2 d_{\rm reg}(F^{\rm top}) - 1$} and \textcolor{black}{$D_1 > 2 d_{\rm reg}(F^{\rm top}) - 1$} occur depending on parameters; the former case tend to hold as $m$ is larger than $n$.}

\textcolor{black}{Salizzoni also proved in \cite{S23} that the solving degree of {\it mutant algorithms} (tamed in \cite{GG23}) such as MutantXL~\cite{buchmann2009mutantxl} and MXL2~\cite{mohamed2008mxl2} is upper-bounded by $d_{\rm reg}(F^{\rm top})  + 1$, but this is not the case that we consider in this paper, since we will construct our algorithm based on the original XL~\cite{CKP00}, not on mutant algorithms.}

\subsection{Complexity}\label{sub:xl_comp}
%==================================

In this subsection, we estimate the time complexity of (plain) XL together with that of its variant Wiedemann XL (WXL).
Here WXL uses Wiedemann's algorithm~\cite{W86} instead of Gaussian elimination in the XL framework, which was first analyzed in~\cite{YCB07}.
Wiedemann's algorithm generally solves sparse linear systems more efficiently than Gaussian elimination.

%=======================
\paragraph{Complexity of XL.}
%=======================
We first consider plain XL (Algorithm \ref{alg:XL}), where the \textbf{Linearize} step is clearly dominant in terms of the time complexity.
Recall from \textcolor{black}{Subsection \ref{subsec:DB}} that XL \textcolor{black}{could output} a solution of the input system for $D$ equal to or larger than \textcolor{black}{$D_0$} given in \eqref{eq:dreg}, and \textcolor{black}{here we assume to take $D$ to be this bound $D_0$}.
In the \textbf{Linearize} step, one uses linear algebra to obtain the reduced row echelon form of a Macaulay matrix with $m \cdot \tbinom{n+D-2}{D-2}$ rows and $\tbinom{n+D}{D}$ columns.
However, in fact, the cost of this step can be estimated as that of Gaussian elimination on a matrix with $\tbinom{n+D}{D}$ rows and columns, assuming the following practical \textcolor{black}{heuristic} as in \cite{Mou11}:
\begin{heur}\hspace{-5pt}{\bf{.}}\label{heur:choose}
    \textcolor{black}{In XL, if} we pick rows \textcolor{black}{in $\mathcal{M}(\mathscr{S}_{\leq D},\mathscr{T}_{\leq D})$} at random under the constraint that we have enough equations at each degree $d \le D$, then usually we have a linearly independent set.
\end{heur}
From this \textcolor{black}{heuristic}, 
%by regarding the complexity of Gaussian elimination as that of LU-decomposition,
the complexity of XL is roughly estimated as
\begin{equation}
	\label{eq:XL}
	O \left( \tbinom{n+D}{D} ^\omega \right),
\end{equation}
where $2 \le \omega < 3$ is \textcolor{black}{the exponent of matrix multiplication}.
%the constant in the complexity of matrix multiplication.

%========================
\paragraph{Complexity of WXL.}
%========================
According to~\cite{UOV}, the complexity of WXL is estimated as
\begin{equation}
	\label{eq:WXL}
	O \left( \tbinom{n}{2} \cdot \tbinom{n+D}{D} ^2 \right),
\end{equation}
where $D$ \textcolor{black}{can be taken to be $D_0$} given in \eqref{eq:dreg}.
(We remove the constant part from the complexity in~\cite{UOV}, since we focus on asymptotic complexity.)
WXL consumes less memory than the plain XL, since it can deal with the Macaulay matrix as a sparse matrix,
and its memory consumption is estimated as $O \left( \tbinom{n}{2} \cdot \tbinom{n+D}{D} \right)$, see~\cite{W86} \textcolor{black}{for details}.

%======================================
\subsection{Improving XL via hybrid approach}\label{sub:hybrid}
%======================================

One of the most effective improvements of XL \textcolor{black}{(Algorithm~\ref{alg:XL})} is to apply the {\it hybrid approach}~\cite{BFP09,YCC04}, which is the best known technique for solving the MQ problem.
The hybrid approach combines an exhaustive search with an MQ solver, and it was proposed in \cite{BFP09} (resp.\ \cite{YCC04}) for Gr\"{o}bner basis algorithms such as $F_4$ and $F_5$ (resp.\ XL).
Specifically, given an MQ system of $m$ equations in $n$ variables, the values for $k$ $(0 \le k \le n)$ variables are randomly guessed and fixed before an MQ solver is applied to the system in the remaining $n-k$ variables; this is repeated until a solution is obtained.
The hybrid approach for XL presented in \cite{YCC04} is called FXL, where ``F'' stands for ``fix'', and it is constructed by adding the first and last steps below into Algorithm~\ref{alg:XL}:
\begin{alg}[Hybrid approach with XL (h-XL)]\hspace{-5pt}{\bf{.}}\label{alg:hXL} 
	~
	\begin{description}
	\item[{\it Input:}] \textcolor{black}{A sequence} $F= ( f_1, \ldots , f_m ) \in \mathbb{F}_q[x_1, \ldots, x_n]^m$ \textcolor{black}{of (not necessarily homogeneous) quadratic polynomials}, the number $k$ of guessed variables, and a degree bound~$D$.
	\item[{\it Output:}] A solution over $\mathbb{F}_q$ to $f_i(x_1, \ldots , x_n) = 0$ for $1 \leq i \leq m$.
	\end{description}
	\begin{enumerate}
		\item {\textbf{Fix:}} Fix the values $a_1,\ldots,a_k \in \mathbb{F}_q$ for the $k$ variables $x_1, \dots, x_k$ randomly.
        \textcolor{black}{In the following two steps, we set $f_i^{({\bf a})}:= f_i(a_1,\ldots,a_k,x_{k+1},\ldots,x_n)$ and $F^{({\bf a})} := (f_1^{({\bf a})},\ldots,f_m^{({\bf a})} )$ with ${\bf a} = (a_1,\ldots,a_k)$.}
		\item \textbf{Multiply:} Construct the shift \textcolor{black}{$\mathscr{S}_{\le D}^{(k)}(F^{({\bf a})}) := \mathscr{T}_{\leq D-2}^{(k)} \cdot F^{({\bf a})}$, where we set $\mathscr{T}_{\leq D-2}^{(k)} := \mathscr{T}(X^{(k)})_{\leq D-2}$ with $X^{(k)} = \{ x_{k+1},\ldots,x_n \}$}.
		\item \textbf{Linearize:} Compute the reduced row echelon form of \textcolor{black}{$\mathcal{M}(\mathscr{S}_{\leq D}^{(k)}(F^{({\bf a})}), \mathscr{T}_{\leq D}^{(k)})$, where we set $\mathscr{T}_{\leq D}^{(k)} := \mathscr{T}(X^{(k)})_{\leq D}$}.
		\item \textbf{Solve:} Compute the root of a univariate polynomial obtained in \textbf{Linearize}.
		\item \textbf{Repeat:} Find the values of the other variables.
		\item If there exists no solution, return to (1) \textbf{Fix}.
	\end{enumerate}
\end{alg}
% \begin{enumerate}
% 	\item[(0)] \textbf{Fix:} Fix the values for the $k$ variables $x_1, \dots, x_k$.
% \end{enumerate}
The complexities of the hybrid approaches using the plain XL and WXL as MQ solvers are estimated as
\begin{eqnarray}
	&&O  \left( q^k \cdot \tbinom{n-k+D}{D} ^\omega \right), \label{eq:hyb_XL} \\
	&&O  \left( q^k \cdot \tbinom{n-k}{2} \cdot \tbinom{n-k+D}{D} ^2 \right), \label{eq:hyb_WXL}
\end{eqnarray}
respectively, by using the estimations \eqref{eq:XL} and \eqref{eq:WXL}.
Here $D$ \textcolor{black}{can be taken as}
\begin{equation}\label{eq:d0nk}
	D_0^{(n-k)}:= \min \left\{ \; d \; \left| \;  \mathrm{coeff}\left( \left( 1-t \right)^{m-(n-k)-1} \left( 1+t \right)^m ,t^d \right) \le d \right. \right\}
\end{equation}
from \eqref{eq:dreg}.
In the use of the hybrid approach, the number $k$ of guessed variables is chosen such that the function inside brackets in \eqref{eq:hyb_XL} or \eqref{eq:hyb_WXL} takes the minimum value.

%=========================================================
\subsection{Crossbred Algorithm}\label{sub:CB}
%=========================================================

This subsection recalls the Crossbred algorithm proposed by Joux and Vitse\textcolor{black}{, which} is a practical efficient algorithm for solving MQ systems over the binary field~\cite{JV17}.
Our proposed algorithm described in Section~\ref{sec:main} \textcolor{black}{follows} a \textcolor{black}{framework similar} to the Crossbred algorithm.
Note that we here change the notation of Crossbred such that it fixes the values of $k$ variables randomly for consistency with the description of our proposed algorithm.

We here roughly describe the Crossbred algorithm.
The Crossbred algorithm takes the number $k$ of guessed variables and the degrees $d$ and $D$ with $d \le D$ as parameters.
In this Crossbred \textcolor{black}{algorithm}, we perform some linear algebra operations on Macaulay matrices before fixing the values of the $k$ variables as in h-XL.
More specifically, for \textcolor{black}{a} given MQ system \textcolor{black}{$F\in \mathbb{F}_{2^r}[x_1,\ldots,x_n]^m$, the Crossbred algorithm} can be described by the following two steps:
The first step generates the Macaulay matrix of the shift of $F$ with degree $\le D$\textcolor{black}{, and then} \textcolor{black}{by linear} algebra on the Macaulay matrix obtains \textcolor{black}{a sequence} $P=(p_1,\dots,p_r)$ \textcolor{black}{of some polynomials whose degrees in the remaining $n-k$ variables are} lower than or equal to $d$.
The second step then performs linear algebra on the Macaulay matrix of the shift of the polynomials obtained by fixing the value of $k$ variables in $F$ and $P$ with degree $\le d$.
If the second step obtains a univariate polynomial, then one can find a solution as in the plain XL algorithm.
This second step is iterated $O(q^k)$ times until one solution is found.

In Subsection~\ref{sub:time_comp} below, we estimate the complexity of \textcolor{black}{the Crossbred algorithm} by Multivariate Quadratic Estimator by the Technology Innovation Institute~\cite{estimator,MQestimator}.
\textcolor{black}{We refer to} \cite{BMS22,Dua23,Nak23} \textcolor{black}{for details on} the complexity of \textcolor{black}{the Crossbred algorithm}.
% \begin{equation}
%     O(n_{cols}^2 + q^{k} \cdot \tilde{n}_cols ^\omega)
% \end{equation}

%================================
\section{Main Algorithm}\label{sec:main}
%================================
In this section, we propose a new variant of the XL algorithm \textcolor{black}{for} solving the MQ problem of $m$ equations in $n$ variables over $\mathbb{F}_q$\textcolor{black}{, in the case} where $n \le m$.
We first discuss Macaulay matrices over polynomial rings, and second describe the outline of our proposed algorithm ``polynomial XL (PXL)".
After that, \textcolor{black}{details} of the most technical step will be described in Subsection~\ref{sub:detail}, and degree bounds for the success of PXL will be discussed in Subsection~\ref{sub:sol_deg}.
\textcolor{black}{
Furthermore, Subsection \ref{subsec:rel} explains the relationship of PXL with FXL and Crossbred,
and Subsection \ref{subsec:toy} gives a toy example.
}
Throughout this section, let $F=(f_1,\dots,f_m)\in {\mathbb F}_q[x_1,\dots, x_n]^m$ be a \textcolor{black}{sequence} of $m$ \textcolor{black}{quadratic (and not necessarily homogeneous)} polynomials in $n$ variables $x_1,\dots,x_n$ over ${\mathbb F}_q$, where $q$ is a power of a prime.

%=========================================================
\subsection{Macaulay matrices over polynomial rings}\label{sub:PMM}
%=========================================================
In this subsection, we fix the notations that are used in the rest of this section.
In particular, we construct a Macaulay matrix {\it over the polynomial ring} $\mathbb{F}_q[x_1, \ldots , x_k]$ with respect to $x_{k+1}, \ldots , x_n$ for $1 \leq k \leq n$, where each entry belongs to $\mathbb{F}_q[x_1, \ldots , x_k]$.
Namely, a Macaulay matrix whose coefficient ring is $\mathbb{F}_q[x_1, \ldots , x_k]$ will be constructed.
Such a Macaulay matrix, together with our construction, plays a key role in the main algorithm in Subsection \ref{sub:out} below.
Note that most of the notations given below are similar to those defined in Subsection~\ref{sub:notation} for the case where the coefficient ring is a general ring.

In the following, an integer $k$ is fixed, unless otherwise noted.
Similarly to the hybrid approach reviewed in Subsection \ref{sub:hybrid}, the main algorithm divides $x_1,\dots,x_n$ into $k$ variables $x_1,\dots,x_k$ and the remaining $n-k$ variables \textcolor{black}{$x_{k+1},\ldots, x_n$}, and then regards $f_1,\dots,f_m$ as elements of the polynomial ring $({\mathbb F}_q[x_1,\dots,x_k])[x_{k+1},\dots, x_n]$.
\textcolor{black}{
As in Subsection~\ref{sub:notation}, we define subsets $\mathscr{T}_d^{(k)}$, $\mathscr{T}_{d';d}^{(k)}$, $\mathscr{T}_{\leq d}^{(k)}$, $\mathscr{S}_{d}^{(k)}$, $\mathscr{S}_{d';d}^{(k)}$, and $\mathscr{S}_{\le d}^{(k)}$ of $(\mathbb{F}_q[x_1, \ldots , x_k])$ $[x_{k+1}, \ldots , x_n]$ as follows:
Putting $X^{(k)} = \{ x_{k+1},\ldots, x_n\}$, we set}
\textcolor{black}{
\begin{align*}
	\mathscr{T}_d^{(k)}  & := \mathscr{T}(X^{(k)})_d = \left\{ x_{k+1}^{\alpha_{k+1}} \cdots x_n^{\alpha_{n}} \in \mathscr{T}(X^{(k)}): \sum_{i=k+1}^{n} \alpha_i = d \right\}, \\
	\mathscr{T}_{d';d}^{(k)} &:= \mathscr{T}_{d'}^{(k)} \cup \mathscr{T}_{d'+1}^{(k)}\cup \cdots \cup \mathscr{T}_{d}^{(k)}, \qquad \mathscr{T}_{\leq d}^{(k)}  :=  \mathscr{T}_{0;d}^{(k)} = \mathscr{T}(X^{(k)})_{\leq d}
\end{align*}
for $0 \leq d' \leq d$, and
\begin{eqnarray*}
	\mathscr{S}_{d}^{(k)} & := & \bigcup_{1\leq i \leq d} \mathscr{T}(X^{(k)})_{d-2}\cdot \{f_i \} =  \{  t f_i : 1 \leq i \leq m,\ t \in \mathscr{T}(X^{(k)})_{d-2}  \}
 \end{eqnarray*}
for $2 \leq d$.
We also set $\mathscr{S}_{0}^{(k)}: = \{ 0 \}$, $\mathscr{S}_{1}^{(k)}: = \{ 0 \}$, and
 \begin{eqnarray*}
% &&\bigcup_{i=1}^m \{ t \cdot f_i : t \in T_{a-2} \}, \\
	\mathscr{S}_{d';d}^{(k)} &:=& \mathscr{S}_{d'}^{(k)} \cup \mathscr{S}_{d'+1}^{(k)} \cup \cdots \cup \mathscr{S}_{d}^{(k)},\qquad \mathscr{S}_{\le d}^{(k)} :=  \mathscr{S}_{0;d}
\end{eqnarray*}
for $0 \leq d' \leq d$.
In particular, $\mathscr{S}_{\leq d}^{(k)}$ is the shift of $F$ by the set $\mathscr{T}_{\leq d-2}^{(k)}$ of monomials in $x_{k+1}, \ldots , x_n$ of degree $\leq d-2$.}

Here, we construct a Macaulay matrix of the shift \textcolor{black}{$\mathscr{S}_{\le D}^{(k)}$} with respect to \textcolor{black}{$\mathscr{T}_{\leq D}^{(k)}$} for $D \geq 2$, as in the plain XL.
For this, unlike the plain XL {(\textcolor{black}{mainly} adopting an elimination order described in Section \ref{sec:pre})}, we use a {\it graded} monomial order (e.g., graded lexicographic order), which is a monomial order first comparing the total degree of two monomials.
Furthermore, as for the order of elements in \textcolor{black}{$\mathscr{S}_{\le D}^{(k)}$}, we also use an order that first compares the degree of two polynomials.
%(As for the order of elements in $I_{\le D}$, we use a typical order described at the end of Subsection \ref{sub:xl}.)
% but in typical for $t,t' \in T_{\leq D}$ we sort $t f_i$ before $t f_j$ if $i < j$, and $t f_i$ before $t' f_i$ if $t \prec t'$, namely
%\[
% I_{\leq D} =  \{ t_1 f_1, t_1 f_2, \ldots , t_1 f_m, t_2 f_1, \ldots , t_{\ell-1} f_m, f_1, \ldots , f_m \}
% \]
% when we write $T_{\leq D} = \{ t_1, \ldots , t_{\ell-1}, 1\}$.

\textcolor{black}{To simplify the} notation, \textcolor{black}{once $F$, $k$, and $D$ are fixed,} we denote the Macaulay matrix \textcolor{black}{$\mathcal{M}(\mathscr{S}_{\leq D}^{(k)},\mathscr{T}_{\leq D}^{(k)})$} constructed as above by $\mathcal{PM}$ \textcolor{black}{to emphasize that it is a {\it polynomial matrix}}, and call it a {\it Macaulay matrix of $F$ at degree $D$ over ${\mathbb F}_q[x_1,\dots, x_k]$}.
For two integers $d_1$ and $d_2$ \textcolor{black}{with $2 \le d_1 \le D$ and $ 0 \le d_2 \le D$}, we also denote by \textcolor{black}{$\mathcal{PM}[\mathscr{S}_{d_1}^{(k)},\mathscr{T}_{d_2}^{(k)}]$} the submatrix of $\mathcal{PM}$ whose rows (resp.\ columns) correspond to polynomials of \textcolor{black}{$\mathscr{S}_{d_1}^{(k)}$ (resp.\ monomials of $\mathscr{T}_{d_2}^{(k)}$)}.
Then, $\mathcal{PM}$ is divided by submatrices \textcolor{black}{$\mathcal{PM}[\mathscr{S}_{d_1}^{(k)},\mathscr{T}_{d_2}^{(k)}]$ for $2 \le d_1 \le D$ and $0 \le d_2 \le D$}.

Thanks to our choice of a graded monomial order together with the quadraticity of $F$, the following lemma \textcolor{black}{holds:}
\begin{lemma}
	\label{lem:PMM}
	For a \textcolor{black}{sequence $F = (f_1,\ldots, f_m)$} \textcolor{black}{of quadratic and not necessarily homogeneous polynomials in $\mathbb{F}_q[x_1,\ldots,x_n]$ and for} positive integers \textcolor{black}{$k$ and $D$ with} \textcolor{black}{$1\leq k \le n$} and $D \ge 2$,
	let $\mathcal{PM}$ be a Macaulay matrix of $F$ at degree $D$ over ${\mathbb F}_q[x_1,\dots, x_k]$.
	Then, for \textcolor{black}{each integer $d$ with} $2 \le d \le D$, \textcolor{black}{the submatrix} $\mathcal{PM}[\mathscr{S}_{d}^{(k)},\mathscr{T}_{d'}^{(k)}]$ with $d' \notin \{d, d-1, d-2\}$ is a zero matrix, and
	all elements of $\mathcal{PM}[\mathscr{S}_d^{(k)},\mathscr{T}_d^{(k)}]$ belong to ${\mathbb F}_q$.
\end{lemma}
\begin{proof}
\textcolor{black}{
Each $f_i$ is written as
\begin{equation}\label{eq:fi}
f_i = q_i(x_{k+1},\ldots,x_n) + \sum_{j=k+1}^n \ell_{i,j}(x_1,\ldots,x_k) x_j + c_i(x_1,\ldots,x_k)
\end{equation}
for a quadratic form $q_i(x_{k+1},\ldots,x_n)$ in $\mathbb{F}_q[x_{k+1},\ldots,x_n]$, linear polynomials $\ell_{i,j}(x_1,\ldots,x_k)$'s in $\mathbb{F}_q[x_{1},\ldots,x_k]$, and a quadratic polynomial $c_i(x_1,\ldots,x_k)$ in $\mathbb{F}_q[x_{1},\ldots,x_k]$.
Therefore, multiplying it by a monomial $t \in \mathscr{T}_{d-2}^{(k)}$ in $x_{k+1},\ldots, x_n$ of degree $d-2$, we have
\[
t f_i = t q_i(x_{k+1},\ldots,x_n) + \sum_{j=k+1}^n \ell_{i,j}(x_1,\ldots,x_k) t x_j + c_i(x_1,\ldots,x_k) t,
\]
where $t q_i$ is a form in $\mathbb{F}_q[x_{k+1},\ldots,x_n]$ of degree $d$ and where each $t x_j$ is a monomial in $\mathbb{F}_q[x_{k+1},\ldots,x_n]$ of degree $d-1$.
This expression of the shift $t f_i$, which corresponds to a row of $\mathcal{PM}[\mathscr{S}_d^{(k)},\mathscr{T}_{\leq d}^{(k)}]$ and vice versa, implies the assertions of the lemma.}
	% The first statement \textcolor{black}{follows} from the fact that \textcolor{black}{the monimials of the non-zero-terms of} $t\cdot f_i \in \mathscr{S}_d^{(k)}$ with $t \in \mathscr{T}_{d-2}^{(k)}$ \textcolor{black}{are monomials of degree $d$, $d-1$, or $d-2$}.
	% Furthermore, if the second one does not hold, then the degree of a given polynomial is larger than 2.
 \qed
\end{proof}

Due to this lemma, we can partly perform row reduction on $\mathcal{PM}$, which is a key operation of the proposed algorithm in the next subsection.

%============================================
\subsection{Outline of our algorithm PXL}\label{sub:out}
%============================================
This subsection describes the proposed algorithm polynomial XL (PXL).
As in the h-XL described in Subsection~\ref{sub:hybrid}, PXL first sets the first $k$ variables $x_1, \dots, x_k$ as guessed variables, whereas the main difference between our PXL and h-XL is the following:
While h-XL performs row reduction after substituting actual $k$ values to $x_1, \dots, x_k$, PXL \textit{partly} performs Gaussian elimination \textit{before} fixing $k$ variables.
These manipulations are possible due to our construction of Macaulay matrices over ${\mathbb F}_q[x_1,\dots, x_k]$ described in Lemma~\ref{lem:PMM}.

Here, we give the outline of PXL.
The notations are same as those in Subsection \ref{sub:PMM}.
\begin{alg}[Polynomial XL]\hspace{-5pt}{\bf{.}}\label{alg:PXL}
	~
	\begin{description}
	\item[{\it Input:}] \textcolor{black}{A sequence} $F=(f_1,\dots,f_m) \in {\mathbb F}_q[x_1,\dots, x_n]^m$ \textcolor{black}{of not necessarily homogeneous polynomials of degree $2$}, the number $k$ of guessed variables, and a degree bound~$D$.
	\item[{\it Output:}] A solution over $\mathbb{F}_q$ to $f_i(x_1, \ldots , x_n) = 0$ for $1 \leq i \leq m$.
	\end{description}
	\begin{enumerate}
		\item \textbf{Multiply}: Compute the set \textcolor{black}{$\mathscr{S}_{\le D}^{(k)}$} of all the products $t \cdot f_i$ with \textcolor{black}{$t \in \mathscr{T}_{\le D-2}^{(k)}$}.
		\item \textbf{Linearize(1)}: Generate \textcolor{black}{$\mathcal{PM}:=\mathcal{M}(\mathscr{S}_{\leq D}^{(k)},\mathscr{T}_{\leq D}^{(k)})$}, which is the Macaulay matrix of $F$ at degree $D$ over $\mathbb{F}_q[x_1,\dots,x_k]$, and partly perform Gaussian elimination on it. (The details will be described in Subsection~\ref{sub:detail} \textcolor{black}{below}.)
		\item \textbf{Fix}: Fix randomly the values for the $k$ variables $x_1,\dots,x_k$ in the resulting matrix of {\bf Linearize(1)}.
		\item \textbf{Linearize(2)}: Compute the \textcolor{black}{reduced} row echelon form of the resulting matrix of step 3. 
%		Perform Gaussian elimination on the resulting matrix of step~3.
		\item \textbf{Solve}: If step~4 yields a univariate polynomial, compute its root.
		\item \textbf{Repeat}: Substitute the root, simplify the equations, and then repeat the process to find the values of the other variables.
		\item If there exists no solution, return to (3) \textbf{Fix}.
	\end{enumerate}
\end{alg}
\noindent
Note that the definition of \textcolor{black}{`the resulting matrix of {\bf Linearize(1)}'} is given in the next paragraph.
% and that the last four steps from \textbf{Fix} to \textbf{Repeat} are iterated until a solution is found.

Let us here describe only the first two steps, since the last four steps are executed similarly to h-XL.
The \textbf{Multiply} step generates the shift \textcolor{black}{$\mathscr{S}_{\leq D}^{(k)}$} of $F$ by \textcolor{black}{$\mathscr{T}_{\leq D-2}^{(k)}$}, defined in Subsection \ref{sub:PMM}, by regarding each polynomial as \textcolor{black}{one} in $({\mathbb F}_q[x_1,\dots, x_k])[x_{k+1},\dots, x_n]$.
At the beginning of the \textbf{Linearize(1)} step, $\mathcal{PM}$ is a polynomial matrix with entries in the polynomial ring $\mathbb{F}_q[x_1,\dots,x_k]$, but by Lemma \ref{lem:PMM} it is almost upper-block triangular, and all of its (nearly-)diagonal blocks are matrices with entries in $\mathbb{F}_q$.
By utilizing this property, the \textbf{Linearize(1)} step repeats to transform such a block into the row echelon form and to eliminate entries of its upper blocks.
After the \textbf{Linearize(1)} step, the resulting Macaulay matrix is supposed to be \textcolor{black}{of the} form
$\begin{pmatrix}
	I & \ast \\
	0 & A
\end{pmatrix}$, by interchanging rows (and columns).
Here $I$ is an identity matrix, and $A$ is a matrix over $\mathbb{F}_q[x_1, \ldots , x_k]$.
% \vspace{-10pt}
% \begin{figure}[H]
% %	\includegraphics[natwidth=1182,natheight=1187,width=10cm,height=10cm]{figure_linearize1.png}
% 	\includegraphics[width=0.5\textwidth,natwidth=900,natheight=270]{figure_linearize1.png}
% 	\centering
% \end{figure}
% \vspace{-10pt}
% \noindent
% with the permutation of rows and columns.
% (All elements in the white parts are zero, whereas all elements in the gray parts and on the diagonal line can be nonzero.)
Then, the last four steps deal with only the submatrix composed of rows and columns including no leading coefficient of the reduced part, which corresponds to $A$.
\textcolor{black}{We call this submatrix $A$ the {\it resulting matrix of} {\bf Linearize(1)}}.
%see the last paragraph of Subsection~\ref{sub:detail} for the precise definition.

%=====================================================
\subsection{\textcolor{black}{Details of \textbf{Linearize(1)} step}}\label{sub:detail}
%=====================================================
In this subsection, we describe the details of the \textbf{Linearize(1)} step in the proposed algorithm, and show that it works well as row operations on $\mathcal{PM}$.
We use the same notations as in Subsection~\ref{sub:PMM}.
In the following, we also denote by \textcolor{black}{$\mathcal{PM}[\mathscr{S}_d^{(k)},\mathscr{T}_{d'}^{(k)}]$} the same part even after $\mathcal{PM}$ is transformed.

The \textbf{Linearize(1)} step is mainly performed on each \textcolor{black}{$\mathcal{PM}[\mathscr{S}_d^{(k)},\mathscr{T}_{(d-2);d}^{(k)}]$}, starting from $d=D$ down to~$2$.
Each iteration $d$ consists of the following three substeps:
\begin{enumerate}
	\setlength{\leftskip}{5mm}
	\item[($d$)-1.] Perform Gaussian elimination on \textcolor{black}{$\mathcal{PM}[\mathscr{S}_d^{(k)},\mathscr{T}_d^{(k)}]$}.
	\item[($d$)-2.] Perform the same row operations as those of ($d$)-1 on the submatrix \textcolor{black}{$\mathcal{PM}[\mathscr{S}_d^{(k)},\mathscr{T}_{(d-2); (d-1)}^{(k)}]$}.
	\item[($d$)-3.] Using the \textit{leading coefficients} of the resulting \textcolor{black}{$\mathcal{PM}[\mathscr{S}_d^{(k)},\mathscr{T}_d^{(k)}]$ (namely the reduced row echelon form of the initial $\mathcal{PM}[\mathscr{S}_d^{(k)},\mathscr{T}_d^{(k)}]$)}, eliminate the corresponding columns of $\mathcal{PM}$. Here, a leading coefficient is the leftmost nonzero entry in each row of \textcolor{black}{a} row echelon form of a matrix.
\end{enumerate}

Here, we show that the \textbf{Linearize(1)} step described above works well as row operations on $\mathcal{PM}$.
Note that for any $3 \le d \le D$, the ($d$)-3 step does not affect the submatrix \textcolor{black}{$\mathcal{PM}[\mathscr{S}_{\le (d-1)}^{(k)},\mathscr{T}_{d}^{(k)}]$, since $\mathcal{PM}[\mathscr{S}_{\le (d-1)}^{(k)},\mathscr{T}_{d}^{(k)}]$ is always a zero matrix by Lemma \ref{lem:PMM}.
This indicates that \textcolor{black}{$\mathcal{PM}[\mathscr{S}_d^{(k)},\mathscr{T}_{\le D}^{(k)}]$}} does not change from the original structure at the beginning of the ($d$)-1 step.
Therefore, from Lemma~\ref{lem:PMM}, the manipulations in the ($d$)-1 and ($d$)-2 steps can be performed correctly and seen as row operations on $\mathcal{PM}$.
Furthermore, the ($d$)-3 step can be also performed correctly, since the leading coefficients of the resulting \textcolor{black}{$\mathcal{PM}[\mathscr{S}_d^{(k)},\mathscr{T}_d^{(k)}]$} belong to $\mathbb{F}_q$.
As a result, we have that all the manipulations are practicable and regarded as row operations on \textcolor{black}{the whole} $\mathcal{PM}$.

After the \textbf{Linearize(1)} step, all manipulations are performed on the resulting matrix of \textbf{Linearize(1)} obtained by concatenating rows and columns including no leading coefficient of the row echelon form $\mathcal{PM}[\mathscr{S}_d^{(k)},\mathscr{T}_d^{(k)}]$ with $2 \le d \le D$.
%In the following, we call this submatrix the resulting matrix of \textbf{Linearize(1)}.

\begin{remark}
\label{rem:row_mac}
%	This remark states that PXL does not need to use all row vectors of Macaulay matrices practically.
%	As in the estimation in Subsection~\ref{sub:xl},
%	we use the following assumption about the number of rows of Macaulay matrices:
%	If we pick rows at random under the constraint that we have enough equations at each degree $d \le D$, then usually we have a linearly independent set.
	As in the XL algorithm, in practice, PXL randomly chooses approximately \textcolor{black}{$|\mathscr{T}_{\le D}^{(k)}|$} independent rows from the Macaulay matrix with \textcolor{black}{$|\mathscr{S}_{\le D}^{(k)}|$} rows \textcolor{black}{(namely we suppose a \textcolor{black}{heuristic} similar to \textcolor{black}{Heuristic \ref{heur:choose}})}, and executes the \textbf{Linearize(1)} step on the submatrix composed of chosen row vectors.
	We then assume that the rank of the resulting matrix of \textbf{Linearize(1)} is large enough to yield a univariate equation, and \textcolor{black}{we experimentally confirmed that this assumption is correct in most cases.}
\end{remark}

%==================================================
\subsection{Degree bounds for the success of PXL}\label{sub:sol_deg}
%==================================================

This subsection estimates the minimum value \textcolor{black}{$D_{\rm PXL}$} where PXL \textcolor{black}{with input $D = D_{\rm PXL}$} succeeds in finding a solution, under \textcolor{black}{a practical assumption (Assumption \ref{assump:semireg_eval} below), which requires conditions similar to (i) and (ii) in Remark~\ref{rem:gradedorder}.}
% the same assumption as in \textcolor{black}{Theorem~\ref{th:KY}, namely we suppose that the sequence $F = (f_1,\ldots,f_m)$ of input (non-homogeneous) polynomials is semi-regular}.
% We call this minimum value \textcolor{black}{$D_{\rm PXL}$} the \textit{solving degree of PXL}.
% \textcolor{black}{
% Specifically, we show that the solving degree of PXL can be upper bounded by the degree of regularity for XL on systems of $m$ equations in $n-k$ variables (cf.\ \eqref{eq:regularity}).}
Note that the success of PXL means the following:
For some evaluation of \textcolor{black}{$\mathbf{a} = (a_1, \ldots , a_k) \in \mathbb{F}_q^k$ to $(x_1, \ldots , x_k)$} in the {\bf Fix} step, the remaining steps finds a solution $(a_{k+1},\ldots , a_n) \in \mathbb{F}_q^{n-k}$ to the multivariate system in $x_{k+1}, \ldots , x_n$ corresponding to the resulting matrix of the {\bf Linearize(1)} step, and then $(a_1, \ldots , a_n)$ is exactly a solution to the original system.

To estimate the \textcolor{black}{value of $D_{\rm PXL}$}, we discuss the rank of the resulting matrix of \textbf{Linearize(1)}.
%derive a sufficient condition for the success of PXL.
Recall from Subsection \ref{sub:out} that the \textbf{Linearize(1)} step  transforms the Macaulay matrix into a matrix of the form $\begin{pmatrix}
I & \ast \\
0 & A
\end{pmatrix}$, by interchanging rows (and columns).
Here $I$ is an identity matrix, and $A$ is a matrix over $\mathbb{F}_q[x_1, \ldots , x_k]$.
The resulting matrix of the {\bf Linearize(1)} step is $A$, and let $\alpha$ be the number of columns of $A$.
For $\mathbf{a} = (a_1, \ldots , a_k) \in \mathbb{F}_q^k$, we denote by $A^{(\mathbf{a})}$ (resp.\ \textcolor{black}{$\mathcal{M}(\mathscr{S}_{\leq D}^{(k)}, \mathscr{T}_{\leq D}^{(k)})^{(\mathbf{a})}$}) the matrix obtained by substituting $(a_1, \ldots , a_k)$ to $(x_1, \ldots , x_k)$ in $A$ (resp.\ \textcolor{black}{$\mathcal{M}(\mathscr{S}_{\leq D}^{(k)}, \mathscr{T}_{\leq D}^{(k)})$}).
Since an evaluation of $x_1, \ldots , x_k$ and elementary row operations over $\mathbb{F}_q[x_1,\dots,x_k]$ (without multiplying rows by elements in $\mathbb{F}_q[x_1,\dots,x_k]$ of degree $\geq 1$) are commutative, we have the following:
\begin{lemma}
\label{lem:rank}
	With notation as above, we have
	\begin{equation*}
		\alpha - \mathrm{rank}(A^{(\mathbf{a})})=|\mathscr{T}_{\leq D}^{(k)}| -\mathrm{rank}(\mathcal{M}(\mathscr{S}_{\leq D}^{(k)}, \mathscr{T}_{\leq D}^{(k)})^{(\mathbf{a})}).
	\end{equation*}
\end{lemma}
Furthermore, we \textcolor{black}{also suppose} the following \textcolor{black}{assumption, in order} to estimate \textcolor{black}{the value of $D_{\rm PXL}$:}
\begin{assump}\hspace{-5pt}{\bf{.}}\label{assump:semireg_eval}
    For any ${\bf{a}}=(a_1, \ldots , a_k) \in \mathbb{F}_q^k$, \textcolor{black}{we have that the sequence $F^{({\bf a})} := (f_1^{({\bf a})},\ldots,f_m^{({\bf a})} )$ with $f_i^{({\bf a})}:= f_i(a_1,\ldots,a_k,x_{k+1},\ldots,x_n)$} satisfies the following conditions.
\begin{enumerate}
        \item[(i)] $F^{({\bf a})}$ has at most one root (counted with multiplicity) over an \textcolor{black}{algebraic} closure $\overline{\mathbb{F}}_q$ of $\mathbb{F}_q$.
        \item[(ii)] $(F^{({\bf a})})^{\rm top}$ is semi-regular (hence it has no root other than $(0,\ldots,0)\! \in \!\mathbb{F}_q^{n-k}\!$).
    \end{enumerate}
\end{assump}
 %satisfies the same assumption about dependencies of $I_{\le d}$ as in Proposition~\ref{pro:deg}.
%This expectation is generally used to estimate the complexity of h-XL and h-WXL, and 
%We experimentally confirmed this statement in most cases.
%Finally, we show how to estimate the solving dgree of PXL.
%
\textcolor{black}{This assumption is expected to hold since $F^{({\bf a})}$ is highly overdetermined (\textcolor{black}{see \cite[Section 4.4]{ikematsu2023recent} for arguments on (ii))}.}
From the above lemma and assumption, we then obtain
\begin{equation*}
    \alpha - \mathrm{rank}(A^{(\mathbf{a})}) = \mathrm{coeff}\left( \left( 1-z \right)^{m-(n-k)-1} \left( 1+z \right)^m ,z^D \right),
\end{equation*}
if $D$ is lower than {$D_{\rm reg}^{(n-k)}$} as in Corollary~\ref{cor:deg}, where
\begin{equation}\label{eq:Dreg_k1}
    D_{\rm reg}^{(n-k)} = \min \left\{ \; d \; \left| \; \mathrm{coeff}\left( \left( 1-z \right)^{m-(n-k)} \left( 1+z \right)^m ,z^d \right) \le 0 \right. \right\}.
\end{equation}
% \textcolor{black}{
% Therefore, as the solving degree of XL is upper-bounded by $D_{reg}$, the solving degree of PXL is also upper-bounded by $D_{reg}$ on systems of $m$ equations in $n-k$ variables given as}
\textcolor{black}{
%Assuming Heuristic \ref{heur:n+1} in the case of $m$ equations in $n-k$ variable, 
Similarly to Remark \ref{rem:gradedorder}, from Assumption \ref{assump:semireg_eval}, we obtain a {\it practical} estimation
\begin{equation}\label{eq:Dreg_k}
  \! \! \! D_{\rm PXL} \!\approx \! D_1^{(n-k)} \!:=\! \min \left\{d \geq 2 \left| \; \! \mathrm{coeff}\! \left( \left( 1-z \right)^{m-(n-k)-1} \! \left( 1+z \right)^m ,z^d \right) \! \le 1 \right. \! \right\}.\!
\end{equation}
Indeed, we experimentally confirmed that PXL finds a solution at $D = D_1^{(n-k)}$.
Note that $D_1^{(n-k)} \geq D_0^{(n-k)}$ for the bound $D_0^{(n-k)}$ given in \eqref{eq:d0nk} for h-XL, but the equality holds in most cases.}
We also note that, as a {\it theoretical} upper-bound on $D_{\rm PXL}$ in the {\it worst} case, we expect from Theorem \ref{th:KY2} that $D_{\rm PXL} \leq 2 D_{\rm reg}^{(n-k)}-1$.

\subsection{Relationships with XFL and Crossbred}\label{subsec:rel}

\begin{remark}[Relationships with XFL~\cite{CKP00,YC04}]
	We here briefly discuss the relationship between our algorithm PXL and XFL~\cite{CKP00,YC04} proposed as a variant of h-XL.
	XFL is roughly described as follows: 
	First, the $k$ variables to be fixed are chosen and generate a shift of the given system by all monomials in the remaining $n-k$ variables up to some degree $D-2$.
	Second, construct a Macaulay matrix (over $\mathbb{F}_q$, but not over $\mathbb{F}_q[x_1,\dots,x_k]$) of the shift with respect to all monomials in the whole $n$ variables up to the degree $D$, and then eliminate only monomials of degree $D$ including only the $n-k$ variables.
%	monomials including only the remaining $n-k$ variables are eliminated before fixing.
	Third, substitute actual values for the $k$ variables, and execute XL for a system in $n-k$ variables obtained by the substitution.

	The first step of XFL clearly coincides with the \textbf{Multiply} step of our PXL.
	The main difference of XFL from PXL is the second step:
	The second step of XFL eliminates monomials in the $n-k$ variables of degree $D$,
	and it corresponds to eliminating only \textcolor{black}{$\mathcal{PM}[\mathscr{S}_D^{(k)},\mathscr{T}_D^{(k)}]$ in the second step of our PXL (in fact, PXL eliminates every block $\mathcal{PM}[\mathscr{S}_d^{(k)},\mathscr{T}_d^{(k)}]$ with $2 \le d \le D$)}.
	%and, if we execute XFL on our Macaulay matrix $\mathcal{PM}$ over $\mathbb{F}_q[x_1,\dots,x_k]$, XFL eliminates only $\mathcal{PM}[I_D,T_D]$.
	%On the other hand, PXL eliminates every block $\mathcal{PM}[I_d,T_d]$ with $2 \le d \le D$.
	Therefore, PXL can be regarded as \textcolor{black}{an} extension of XFL, and the size of the uneliminated part of the second step of XFL is larger than that of PXL.
	
%	XFL is similar to PXL in the point that the equations are partly eliminated before fixing.
%	However, PXL eliminates some monomials not only over the field but also over the polynomial ring of the $k$ variables.
	%
\end{remark}

\begin{remark}[Relationships with Crossbred~\cite{JV17}]
%	This remark discusses the relationship between our PXL and the Crossbred algorithm proposed by Joux and Vitse~\cite{JV17}.
	This remark explains the difference between our PXL and the Crossbred algorithm proposed by Joux and Vitse~\cite{JV17}\textcolor{black}{,} from the following two points:
 \textcolor{black}{(i) The} \textcolor{black}{parts of Macaulay matrices} echelonized before the fixing step\textcolor{black}{, and (ii) Our} original structure of Macaulay matrices over the polynomial ring $\mathbb{F}_q [x_1,\dots,x_k]$\textcolor{black}{,} where $x_1,\dots,x_k$ are \textcolor{black}{variables to be fixed}.

	First, \textcolor{black}{the parts of Macaulay matrices echelonized before the fixing step for PXL are definitely different from those for Crossbred by the following reason:}
	Crossbred eliminates monomials in which the degree of the remaining $n-k$ variables is larger than \textcolor{black}{a} given degree, whereas \textcolor{black}{our algorithm PXL} eliminates \textcolor{black}{$\rank\left(\mathcal{PM}[\mathscr{S}_d^{(k)},\mathscr{T}_{d}^{(k)}]\right)$} monomials among degree $d$ monomials in the $n-k$ variables for each $2 \le d \le D$.
	\textcolor{black}{This could cause a difference in the estimations of the degrees $D$ (for which a root is found) and the complexities}.

	Second, our Macaulay matrix is constructed over $\mathbb{F}_q[x_1,\dots,x_k]$ by regarding each polynomial in $\mathbb{F}_q[x_1,\dots,x_n]$ as an element of the polynomial ring in the $n-k$ variables over $\mathbb{F}_q[x_1,\dots,x_k]$, unlike Crossbred, which uses \textcolor{black}{a} Macaulay matrix over the base field $\mathbb{F}_q$.
	In our Macaulay matrix over $\mathbb{F}_q[x_1,\dots,x_k]$, row operations adding a multiple of one row with one variable $x_i$ with $1 \le i \le k$ into another row can be realized. 
	By contrast, such a row operation cannot be performed in the standard Macaulay matrix over $\mathbb{F}_q$ clearly.
	Therefore, row reductions performed in our PXL cannot be duplicated in the standard Macaulay matrix over $\mathbb{F}_q$, and thus row reductions of our PXL performed before fixing the values of $k$ variables are different from those of Crossbred.

	% Comparing our PXL and Crossbred,
	% submatrices of the Macaulay matrix that we perform Gaussian elimination before guessing the values of some variables are clearly different between them.
	% Furthermore, in a technical point, we utilize the Macaulay matrix over the polynomial ring, unlike Crossbred.

	% In terms of efficiency, to the best of our knowledge, there exists no result that shows the asymptotic efficiency of the Crossbred compared with h-XL and h-WXL.
	% Therefore, we compare the complexity of our PXL only with those of h-XL and h-WXL in Subsection~\ref{sub:time_comp} below.

\end{remark}

\subsection{Toy Example}\label{subsec:toy}

We here solve an MQ system $F=(f_1,f_2,f_3)$ in \textcolor{black}{$n = 3$} variables $(x_1,x_2,x_3)$ over $\mathbb{F}_7$ of \textcolor{black}{$m = 3$ polynomials}
\begin{align*}
	&f_1=5x_1^2+6x_1x_2+4x_1x_3+x_2x_3+5x_3^2+4x_1+5x_2+3 , \\
	&f_2=4x_1^2+5x_1x_2+4x_1x_3+3x_2^2+5x_2x_3+x_3^2+6x_1+2x_2+3x_3+2 , \\
	&f_3=2x_1^2+4x_1x_2+2x_2^2+6x_3^2+6x_1+x_2+3x_3+2 ,
\end{align*}
by our PXL with $k=1$ and $D=4$; in fact, we can take \textcolor{black}{$D=3$} by $D_1^{(n-k)} = 3$ from \eqref{eq:Dreg_k} (or $2 d_{\rm reg}(F^{\rm top})-1 = 3$ in Theorem \ref{th:KY2}), but we take $D=4$ for a demonstration.

\vspace{4pt}
Then the Macaulay matrix $\mathcal{PM}$ \textcolor{black}{of $F$ at degree $D$} over $\mathbb{F}_7[x_1]$ is given as
\scriptsize
\begin{eqnarray*}
    &
    \renewcommand{\kbldelim}{(}
    \renewcommand{\kbrdelim}{.}
    \kbordermatrix{
			& x_2^4 & x_2^3 x_3 & x_2^2 x_3^2 & x_2 x_3^3 & x_3^4 &\vrule
			& x_2^3 & x_2^2 x_3 & x_2 x_3^2 & x_3^3
			 \\
			x_2^2 f_1 &  & 1 & 5 &  &  &\vrule
			& 6x_1+5 & 4x_1 &  &  
			\\
			x_2^2 f_2 & 3 & 5 & 1 &  &  &\vrule
			& 5x_1+2 & 4x_1+3 &  &  
			\\
			x_2^2 f_3 & 2 &  & 6 &  &  &\vrule
			& 4x_1+1 & 3 &  &  
			\\
			x_2x_3 f_1 &  &  & 1 & 5 &  &\vrule
			&  & 6x_1+5 & 4x_1 &  
			\\
			x_2x_3 f_2 &  & 3 & 5 & 1 &  &\vrule
			&  & 5x_1+2 & 4x_1+3 &  
			\\
			x_2x_3 f_3 &  & 2 &  & 6 &  &\vrule
			&  & 4x_1+1 & 3 &  
			\\
			x_3^2 f_1 &  &  &  & 1 & 5 &\vrule
			&  &  & 6x_1+5 & 4x_1 
			\\
			x_3^2 f_2 &  &  & 3 & 5 & 1 &\vrule
			&  &  & 5x_1+2 & 4x_1+3 
			\\
			x_3^2 f_3 &  &  & 2 &  & 6 &\vrule
			&  &  & 4x_1+1 & 3 
			\\
			%%%
                \cline{1-1}\cline{2-11}
			x_2 f_1 &  &  &  &  & &\vrule &  & 1 & 5 
			\\
			x_2 f_2 &  &  &  &  & &\vrule & 3 & 5 & 1 
			\\
			x_2 f_3 &  &  &  &  & &\vrule & 2 &  & 6 
			\\
			x_3 f_1 &  &  &  &  & &\vrule &  &  & 1 & 5 
			\\
			x_3 f_2 &  &  &  &  & &\vrule &  & 3 & 5 & 1 
			\\
			x_3 f_3 &  &  &  &  & &\vrule &  & 2 &  & 6 
			\\
   \cline{1-1}\cline{2-11}
			f_1 &  &  &  &  & &\vrule &  &  &  &  \\
			f_2 &  &  &  &  & &\vrule &  &  &  &  \\
			f_3 &  &  &  &  & &\vrule &  &  &  &   \\
		} \hspace{100pt}\\
  &\\
  & %\hspace{-50pt}
      \renewcommand{\kbldelim}{.}
    \renewcommand{\kbrdelim}{)}
		\kbordermatrix{
			& {x_2^2} & {x_2 x_3} & {x_3^2} &\vrule & {x_2} & {x_3} &\vrule & {1} \\
			& 5x_1^2+4x_1+3 &  &  & \vrule & &  & \vrule & \\
			& 4x_1^2+6x_1+2 &  &  & \vrule & &  & \vrule & \\
			& 2x_1^2+6x_1+2 &  &  & \vrule & &  & \vrule & \\
			&  & 5x_1^2+4x_1+3 &  & \vrule & &  & \vrule & \\
			&  & 4x_1^2+6x_1+2 &  & \vrule & &  & \vrule & \\
			&  & 2x_1^2+6x_1+2 &  & \vrule & &  & \vrule & \\
			&  &  & 5x_1^2+4x_1+3 & \vrule & &  & \vrule & \\
			&  &  & 4x_1^2+6x_1+2 & \vrule & &  & \vrule & \\
			&  &  & 2x_1^2+6x_1+2 & \vrule & &  & \vrule & \\
			%%% 
   \cline{2-9}
			& 6x_1+5 & 4x_1 &  &\vrule & 5x_1^2+4x_1+3 &  &   \vrule & \\
			& 5x_1+2 & 4x_1+3 &  &\vrule & 4x_1^2+6x_1+2 &  &   \vrule & \\
			& 4x_1+1 & 3 &  &\vrule & 2x_1^2+6x_1+2 &  &   \vrule & \\
			&  & 6x_1+5 & 4x_1 &\vrule &  & 5x_1^2+4x_1+3 & \vrule & \\
			&  & 5x_1+2 & 4x_1+3 &\vrule &  & 4x_1^2+6x_1+2 & \vrule & \\
			&  & 4x_1+1 & 3 &\vrule &  & 2x_1^2+6x_1+2 & \vrule & \\
   \cline{2-9}
			&  & 1 & 5 &\vrule & 6x_1+5 & 4x_1 &\vrule & 5x_1^2+4x_1+3 \\
			& 3 & 5 & 1 &\vrule & 5x_1+2 & 4x_1+3 &\vrule & 4x_1^2+6x_1+2 \\
			& 2 &  & 6 &\vrule & 4x_1+1 & 3 &\vrule & 2x_1^2+6x_1+2 \\
		} ,\\
\end{eqnarray*}
\normalsize
and this can be regarded as a block matrix with the following form:
{\tiny
\begin{equation*}
    \begin{pmatrix}
             ~\mathcal{PM}[\mathscr{S}_4^{(1)},\mathscr{T}_4^{(1)}]~ &\vrule& ~\mathcal{PM}[\mathscr{S}_4^{(1)},\mathscr{T}_3^{(1)}]~ &\vrule& ~\mathcal{PM}[\mathscr{S}_4^{(1)},\mathscr{T}_2^{(1)}]~ &\vrule& ~\mathcal{PM}[\mathscr{S}_4^{(1)},\mathscr{T}_1^{(1)}]~ &\vrule& ~\mathcal{PM}[\mathscr{S}_4^{(1)},\mathscr{T}_0^{(1)}]~\\[3pt]
             \cline{1-9}
             \mathcal{PM}[\mathscr{S}_3^{(1)},\mathscr{T}_4^{(1)}] &\vrule& \mathcal{PM}[\mathscr{S}_3^{(1)},\mathscr{T}_3^{(1)}] &\vrule& \mathcal{PM}[\mathscr{S}_3^{(1)},\mathscr{T}_2^{(1)}] &\vrule& \mathcal{PM}[\mathscr{S}_3^{(1)},\mathscr{T}_1^{(1)}] &\vrule& \mathcal{PM}[\mathscr{S}_3^{(1)},\mathscr{T}_0^{(1)}] \\[3pt]
             \cline{1-9}
             \mathcal{PM}[\mathscr{S}_2^{(1)},\mathscr{T}_4^{(1)}] &\vrule& \mathcal{PM}[\mathscr{S}_2^{(1)},\mathscr{T}_3^{(1)}] &\vrule& \mathcal{PM}[\mathscr{S}_2^{(1)},\mathscr{T}_2^{(1)}] &\vrule& \mathcal{PM}[\mathscr{S}_2^{(1)},\mathscr{T}_1^{(1)}] &\vrule& \mathcal{PM}[\mathscr{S}_2^{(1)},\mathscr{T}_0^{(1)}] \\[3pt]
    \end{pmatrix}.
\end{equation*}
}

%(K[x_1,..,x_n]/I)_1
%
In the {\bf Linearize(1)} step, we first perform the Gaussian elimination on \textcolor{black}{$\mathcal{PM}[\mathscr{S}_4^{(1)},\mathscr{T}_4^{(1)}]$, and then $\mathcal{PM}[\mathscr{S}_4^{(1)},\mathscr{T}_{2;4}^{(1)}]$} is changed into
\tiny
\begin{eqnarray*}
		\kbordermatrix{
			& x_2^4 & x_2^3 x_3 & x_2^2 x_3^2 & x_2 x_3^3 & x_3^4 &\vrule& x_2^3 & x_2^2 x_3 & x_2 x_3^2 & x_3^3
			&\vrule& {x_2^2} & {x_2 x_3} & {x_3^2}  \\
			& 1 &  &  &  &  &\vrule&4 x_1 + 3 & 2 & 6x_1 +3 & 
			&\vrule& 6x_1^2 + 2x_1 +3 & 3x_1^2 + x_1 & \\
            &  & 1 &  &  &      &\vrule& 5x_1 +3  & 3x_1 +2 & 5x_1 +6 & 
			&\vrule& 3x_1^2 + x_1 +6 & 6x_1^2 + 2x_1 &  \\
            &  &  & 1 &  &      &\vrule& 3x_1 +6  & 3x_1 +1 & 6x_1 +3 &   
			&\vrule& 6x_1^2 + 2x_1 +5 & 3x_1^2 + x_1 &  \\
            &  &  &  & 1 &      &\vrule& 5x_1 +3  & 2x_1 +5 & x_1 +5 &   
			&\vrule& 3x_1^2 + x_1 +6 & 6x_1^2 + 2x_1 +2 &  \\
            &  &  &  &  & 1    &\vrule& 6x_1 +5  & x_1 +6 & x_1 & 5x_1  
			&\vrule& 5x_1^2 + 4x_1 +3 & 3x_1^2 + x_1 +1 & x_1^2 + 5x_1 +2 \\
   &  &  &  &  &      &\vrule& 2x_1 +5  & x_1 & 5x_1 +6 &  
			&\vrule& x_1^2 + 6x_1 +5 & 6x_1^2 + 2x_1 &  \\
            &  &  &  &  &      &\vrule& 3x_1 +6  & 3 & 4x_1 +1 &  
			&\vrule& 6x_1^2 + 2x_1 +5 & x_1^2 + 6x_1 +6 &  \\
            &  &  &  &  &      &\vrule&  & 3x_1 +6 & 3 & 4x_1 +1 
			&\vrule&  & 6x_1^2 + 2x_1 +5 & x_1^2 + 6x_1 +6 \\
            &  &  &  &  &      &\vrule& 6x_1 +5  & 3x_1 +3 & 6x_1 +2 & 4x_1 +2 
			&\vrule& 5x_1^2 + 4x_1 +3 & 2 & 2x_1^2 + 3x_1 \\
			%%%
		}.
\end{eqnarray*}
\normalsize
Note that the first five rows of the above matrix can be ignored after this elimination.
We then \textcolor{black}{eliminate} elements of $\mathcal{PM}[\mathscr{S}_3^{(1)},\mathscr{T}_3^{(1)}]$, and then $\mathcal{PM}[\mathscr{S}_3^{(1)},\mathscr{T}_{1;3}^{(1)}]$ is changed into
\begin{eqnarray*}
    \kbordermatrix{
         & x_2^3 & x_2^2 x_3 & x_2 x_3^2 & x_3^3&\vrule
        & {x_2^2} & {x_2 x_3} & {x_3^2} &\vrule& x_2 & x_3 \\
            & 1 &  &  &  &\vrule& 4x_1 +3 & 2 & 6x_1 +3 &\vrule& 6x_1^2 + 2x_1 +3 & 3x_1^2 + x_1 \\
            &  & 1 &  &  &\vrule& 5x_1 +3 & 3x_1 +2 & 5x_1 +6 &\vrule& 3x_1^2 + x_1 +6 & 6x_1^2 + 2x_1 \\
            &  &  & 1 &  &\vrule& 3x_1 +6 & 3x_1 +1 & 6x_1 +3 &\vrule& 6x_1^2 + 2x_1 +5 & 3x_1^2 + x_1 \\
            &  &  &  & 1 &\vrule& 5x_1 +3 & 2x_1 +5 & x_1 +5 &\vrule& 3x_1^2 + x_1 +6 & 6x_1^2 + 2x_1 +2 \\
            &  &  &  &  &\vrule& 3x_1 +6 & 3 & 4x_1 +1 &\vrule& 6x_1^2 + 2x_1 +5 & x_1^2 + 6x_1 +6 \\
            &  &  &  &  &\vrule& 4x_1 +3 & 2x_1 & 3x_1 +5 &\vrule& 2x_1^2 + 5x_1 +3 & 5x_1^2 + 4x_1 \\
    }.
\end{eqnarray*}
Then using the leading \textcolor{black}{coefficient} of \textcolor{black}{this partly reduced $\mathcal{PM}[\mathscr{S}_3^{(1)},\mathscr{T}_3^{(1)}]$}, we eliminate nonzero elements of the last four rows of \textcolor{black}{$\mathcal{PM}[\mathscr{S}_4^{(1)},\mathscr{T}_3^{(1)}]$}, and then the last four rows of \textcolor{black}{$\mathcal{PM}[\mathscr{S}_4^{(1)},\mathscr{T}_{1;2}^{(1)}]$} becomes the following form 
\small
\begin{eqnarray*}
    \kbordermatrix{
        & {x_2^2} & {x_2 x_3} & {x_3^2} &\vrule& x_2 & x_3 \\
        & x_1^2 + 6x_1 +3 & 2x_1^2 + x_1 +5 & 2x_1^2 + 5x_1 +2 &\vrule& 4x_1^3 + 3x_1^2 + 4x_1 +4 & x_1^3 + 3x_1 \\
        & 3x_1^2 + 4x_1 & 3x_1^2 + 5x_1 +1 & 6x_1 +3 &\vrule& 5x_1^2 + 3x_1 +1 & 3x_1^2 + x_1 \\
        & 5x_1 +3 & 3x_1^2 + 3x_1 +6 & 3x_1^2 + 3x_1 +5 &\vrule& 3x_1^2 + x_1 +6 & 5x_1^2 + 3x_1 +5 \\
        & 5x_1^2 + 4x_1 +3 & 2 & 2x_1^2 + 3x_1 &\vrule& 5x_1^3 + 3x_1^2 + 3x_1 +1 & 6x_1^3 + 3x_1 +3 \\
    }.
\end{eqnarray*}
\normalsize
\textcolor{black}{Similarly,} we perform the ($d$)-1, ($d$)-2, and ($d$)-3 steps with $d=2$, and then the resulting matrix of the {\bf Linearize(1)} step is given as
% \begin{eqnarray*}
%     \begin{split}
%         \kbordermatrix{
%             & {x_2^2} & {x_2 x_3} & {x_3^2} & x_2 & x_3 & 1 \\
%             %
%             & 1 &  &  & 3x_1 +4 & 3x_1 +2 & 2x_1^2 + 6x_1 +4 \\
%             &  & 1 &  & 3x_1 +5 & 2x_1 +2 & 2x_1^2 + 2x_1 +1 \\
%             &  &  & 1 & 2x_1 & 6x_1 +1 & 2x_1^2 + 6x_1 +6 \\ 
%         }.
%     \end{split}
% \end{eqnarray*}
\begin{eqnarray*}
%    \begin{split}
        \kbordermatrix{
            & x_2 & x_3 & 1 \\
            & 5x_1^3 + 3x_1 +2 & ~3x_1^3 + 5x_1^2 + 2x_1 +3~ & 4x_1^4 + 3x_1^3 + 6x_1^2 + 3x_1 +6 \\
            & 3x_1^3 + 2x_1^2 + 2x_1 +3 & 6x_1^3 + 3x_1^2 + 6x_1 +2 & 2x_1^4 + 2x_1^3 + 5x_1^2 +2 \\
            & 6x_1^3 + 6x_1 +6 & 4x_1^3 + 6x_1^2 + 3x_1 +3 & 2x_1^4 + 3x_1^3 + 2x_1^2 + 4x_1 +1 \\
            & 3x_1^2 + 3x_1 +1 & 3x_1^2 + x_1 +1 & x_1^2 + 2 \\
            & 6x_1^2 + 2x_1 +5 & 6x_1^2 + 6x_1 +3 & 3x_1^3 + x_1^2 \\
        }.
%    \end{split}
\end{eqnarray*}

In the {\bf Fix} step, we here substitute $x_1=3$ and obtain the following matrix by the Gaussian elimination
% \begin{eqnarray*}
%     \begin{split}
%         \kbordermatrix{
%             & x_2 & x_3 & 1 \\
%             %
%             & 6 & 2 & 5 \\
%             & 3 & 6 & 4 \\
%             & 4 & 6 & 1 \\
%             & 4 & 2 & 4 \\
%             & 2 & 5 & 6 \\
%         }.
%     \end{split}
% \end{eqnarray*}
\begin{eqnarray*}
  %  \begin{split}
        \kbordermatrix{
            & x_2 & x_3 & 1 \\
            & 1 &  & 4 \\
            &  & 1 & 1 \\
            % &  &  &  \\
            % &  &  &  \\
            % &  &  &  \\
        }.
   % \end{split}
\end{eqnarray*}
Then, we can obtain two univariate equations $x_2+4=0$ and $x_3+1=0$,
and thus a solution is $(x_1,x_2,x_3)=(3,3,6)$.

%================
\section{Complexity}\label{sec:complexity}
%================
In this section, we first estimate the size of the resulting matrix of \textbf{Linearize(1)}.
After that, we estimate the time complexity of PXL and compare it with those of h-XL, h-WXL, and Crossbred.
\textcolor{black}{We take $D$ to be $D_{1}^{(n-k)}$} so that PXL can find a solution (as described in Subsection \ref{sub:sol_deg}).

%==============================================================
\subsection{Size of resulting matrix of \textbf{Linearize(1)}}\label{sub:alpha}
%==============================================================

Let $\alpha$ be the number of columns of the resulting matrix of \textbf{Linearize(1)}.
In the following, we estimate the value of this $\alpha$,
and show that it can be quite smaller than the number of the columns of the original Macaulay matrix $\mathcal{PM}$.
We also \textcolor{black}{describe} that the resulting matrix of \textbf{Linearize(1)} can be assumed to be an $\alpha \times \alpha$ matrix.

%From the discussion in Section~\ref{sec:main}, $PM[A,B]$ is a nearly square matrix, thus we only discuss the size of a index set $B$.
\textcolor{black}{
As in the proof of Lemma \ref{lem:PMM}, we denote by $q_i$ the sum of degree-$2$ terms with respect to $x_{k+1},\ldots,x_n$ in $f_i$.
Note that $q_i = f_i^{\rm top}(0,\ldots,0,x_{k+1},\ldots,x_n)$ for each $i$ with $1 \leq i \leq m$.
Then, by putting $F^{({\rm top},k)} := (q_1,\ldots,q_m)$, it is straightforward that the elements in the shift $\mathscr{S}_d^{(k)}(F^{({\rm top},k)})$, which is equal to $((\mathscr{S}_d^{(k)})^{\rm top})|_{(x_1,\ldots,x_k)=(0,\ldots,0)}$, correspond to the rows of $\mathcal{PM}[\mathscr{S}_d^{(k)},\mathscr{T}_d^{(k)}]$, for each non-negative integer $d$. 
We have that the number of columns eliminated in the step ($d$)-1 of \textbf{Linearize(1)} on $\mathcal{PM}[\mathscr{S}_d^{(k)},\mathscr{T}_d^{(k)}]$ is equal to the rank of $\mathcal{PM}[\mathscr{S}_d^{(k)},\mathscr{T}_d^{(k)}]$, that is $\mathrm{dim}_{\mathbb{F}_q}\langle \mathscr{S}_d^{(k)}(F^{({\rm top},k)})\rangle_{\mathbb{F}_q}$.
Therefore, we have}
\begin{eqnarray}
	\alpha &=& |\mathscr{T}_{\le D}^{(k)}| - \sum_{d=0}^D \mathrm{dim}_{\mathbb{F}_q}\langle \mathscr{S}_d^{(k)}(F^{({\rm top},k)}) \rangle_{\mathbb{F}_q} \nonumber \\
 &=& \sum_{d=0}^D \left( |\mathscr{T}_d^{(k)}|-\mathrm{dim}_{\mathbb{F}_q}\langle \mathscr{S}_d^{(k)}(F^{({\rm top},k)}) \rangle_{\mathbb{F}_q}\right) \nonumber \\
 & = & \sum_{d=0}^D \left( \dim_{\mathbb{F}_q} \mathbb{F}_q[x_{k+1},\ldots,x_n]_d -\mathrm{dim}_{\mathbb{F}_q}\langle F^{({\rm top},k)} \rangle_{d}\right) \nonumber \\
 &=& \sum_{d=0}^D \dim_{\mathbb{F}_q} \mathbb{F}_q[x_{k+1},\ldots,x_n]_d / \langle F^{({\rm top},k)} \rangle_{d},  \label{eq:alpha_def}
\end{eqnarray}
% \begin{eqnarray}
% 	\label{eq:alpha_def}
% 	\alpha &=& |\mathscr{T}_{\le D}^{(k)}| - \sum_{d=0}^D \mathrm{dim}_{\mathbb{F}_q}(\langle I^*_{d} \rangle _{\mathbb{F}_q}) \nonumber 
% 	=& \sum_{d=2}^D \left( |T_d|-\mathrm{dim}_{\mathbb{F}_q}(\langle I^*_{d} \rangle _{\mathbb{F}_q}) \right) + |T_1| + |T_0|.
% \end{eqnarray}
%where $\bar{I}_1$ and $\bar{I}_0$ are set to be emptysets.
%Using the same assumption as in Proposition~\ref{pro:deg}, 
\textcolor{black}{where we used $\langle \mathscr{S}_d^{(k)}(F^{({\rm top},k)}) \rangle_{\mathbb{F}_q} = \langle F^{({\rm top},k)} \rangle_{d}$ since all the elements in the sequence $F^{({\rm top},k)}=(q_1,\ldots, q_m)$ are homogeneous.
Here, we suppose the following:}
\begin{assump}\hspace{-5pt}{\bf{.}}\label{assump:q_semireg}
\textcolor{black}{
    The sequence $F^{({\rm top},k)}=(q_1,\ldots, q_m)$ of homogeneous polynomials in $\mathbb{F}_q[x_{k+1},\ldots,x_n]$ is semi-regular, where $q_i$ is given in \eqref{eq:fi} of the proof of Lemma \ref{lem:PMM}.}
\end{assump}
Under this assumption, the value of~\eqref{eq:alpha_def} can be estimated as
\begin{equation}
\label{eq:alpha}
	\alpha = \sum_{d=0}^D \max \left\{ \mathrm{coeff}\left( \left( 1-t \right)^{m-(n-k)} \left( 1+t \right)^m ,t^d \right) ,0 \right\}
%	\sum_{d=0}^D \max \left\{ \left( [t^d]\{ (1-t)^{m-(n-k)} (1+t)^m \} \right) ,0 \right\}.
%	&&= \max_{1 \le d \le D} \{ [t^d]\{ (1-t)^{m-(n-k)-1} (1+t)^m \} \}.
\end{equation}
%For example, when $n=m=40$ and $k=10$, the solving degree $D$ of PXL obtained by~\eqref{eq:regularity} is 10, and $\tbinom{n+D}{D}$ and $\alpha$ estimated by~\eqref{eq:alpha} are approximately $2^{30}$ and $2^{21}$, respectively.
%
\textcolor{black}{by Proposition \ref{prop:Diem}.}
%This is because $|\mathscr{T}_d^{(k)}| = \dim_{\mathbb{F}_q} \mathbb{F}_q[x_{k+1},\ldots,x_n]_d$ and for .}
Note that this can be quite smaller than \textcolor{black}{$\tbinom{n-k+D}{D}$}, which is the number of the columns of the whole Macaulay matrix $\mathcal{PM}$.
For example, when $n=m=40$ and $k=10$, \textcolor{black}{the degree $D$ for which PXL could succeed is estimated as $10$ by~\eqref{eq:Dreg_k}}, and then $\alpha$ and \textcolor{black}{$\tbinom{n-k+D}{D}$} are approximately $2^{21}$ and $2^{30}$, respectively.

% Furthermore, we discuss the number of rows required for executing PXL practically.
% As in the estimation in Subsection~\ref{sub:xl},
% we use the following assumption about the number of rows of Macaulay matrices:
% If we pick rows at random under the constraint that we have enough equations at each degree $d \le D$, then usually we have a linearly independent set.
% From this assumption, PXL randomly chooses approximately $|T_{\le D}|$ independent row vectors from the Macaulay matrix with $|I_{\le D}|$ rows and executes the \textbf{Linearize(1)} step on the submatrix.

Recall from Remark~\ref{rem:row_mac} that PXL randomly chooses approximately \textcolor{black}{$|\mathscr{T}_{\le D}^{(k)}|$} independent rows from the \textcolor{black}{whole} Macaulay matrix \textcolor{black}{$\mathcal{PM}$}.
When \textcolor{black}{$\tilde{\mathscr{S}}_d^{(k)}$} denotes the subset of \textcolor{black}{$\mathscr{S}_d^{(k)}$} including polynomials corresponding to randomly chosen rows and \textcolor{black}{$r_d^{(k)}$} denotes the rank of \textcolor{black}{$\mathcal{PM}[\tilde{\mathscr{S}}_d^{(k)},\mathscr{T}_d^{(k)}]$, the number of rows of  the resulting matrix of \textbf{Linearize(1)} is equal to $\sum_{d=2}^D \left( |\tilde{\mathscr{S}}_d^{(k)}| -r_d^{(k)} \right)$, and we {\it suppose} the following approximation:
\begin{equation}
	\label{eq:row_mac}
\sum_{d=2}^D \left( |\tilde{\mathscr{S}}_d^{(k)}| -r_d^{(k)} \right) \approx \alpha.
\end{equation}
This can be realized by avoiding choosing too many rows from $\mathscr{S}_D^{(k)}$}, and,
by doing so, the size of the resulting matrix of \textbf{Linearize(1)} is approximately $\alpha \times \alpha$.

% Recall from Remark~\ref{rem:row_mac} that PXL randomly chooses approximately $|T_{\le D}|$ rows from the Macaulay matrix.
% In the following, we suppose that approximately $|T_d|-\mathrm{coeff}\left( \left( 1-t \right)^{m-(n-k)} \left( 1+t \right)^m ,t^d \right)$ rows are chosen from $I_d$ for each $2 \le d \le D-1$.
% By doing so, we can assume that the size of the resulting matrix of \textbf{Linearize(1)} is approximately $\alpha \times \alpha$.
% Furthermore, when $\tilde{I}_d$ denotes the subset of $I_d$ including polynomials corresponding to randomly chosen rows, we have
% \begin{equation}
% \label{eq:row_mac}
% 	\sum_{d=2}^{D}
% 	\max \left\{ |\tilde{I}_d| - |T_d| ,0 \right\}
% 	\approx \alpha.
% \end{equation}

%=========================
\subsection{Time complexity}
%=========================

In this subsection, we estimate the time complexity of PXL.
Here, $C_{(d)1}$ (resp.\ $C_{(d)2}$, $C_{(d)3}$) denotes the estimation of the sum of the number of operations in $\mathbb{F}_q$ required for $(d)-1$ (resp.\ $(d)-2$, $(d)-3$) in the \textbf{Linearize(1)} step \textcolor{black}{for all $d$} with $2 \le d \le D$.
Furthermore, $C_{\mathsf{fix}}$ (resp.\ $C_{\mathsf{li2}}$) denote the estimation of the number of operations in $\mathbb{F}_q$ required for the \textbf{fix} (resp.\ \textbf{Linearize(2)}) step.
These estimations are determined from the number $n$ of all variables, the number $k$ of guessed variables, \textcolor{black}{the degree bound $D$ (which can be taken to be $D_1^{(n-k)}$ given in \eqref{eq:Dreg_k})}, and the size $\alpha$ of the resulting matrix of \textbf{Linearize(1)}.
After obtaining each of these five estimations,
we give a practical estimation of total time complexity by~\eqref{eq:pro_odd} below.

%=================================
\paragraph*{Time Complexity of $(d)$-1.}
%=================================	
Recall that the $(d)$-1 step performs Gaussian elimination on \textcolor{black}{$\mathcal{PM}[\tilde{\mathscr{S}}_d^{(k)},\mathscr{T}_d^{(k)}]$}, and its complexity is given as \textcolor{black}{$\max \{|\tilde{\mathscr{S}}_d^{(k)}|, |\mathscr{T}_d^{(k)}| \}^\omega$} for each \textcolor{black}{$d$ with} $2 \le d \le D$.
%Since the complexity of the $(d)$-1 step is $\max \{|\tilde{I}_d|, |T_d| \} ^\omega$ for each $2 \le d \le D$,
Since we have \textcolor{black}{$\sum_{d=2}^D  \max\{|\tilde{\mathscr{S}}_d^{(k)}| -|\mathscr{T}_d^{(k)}|, 0\}  \le \alpha$} from \eqref{eq:row_mac}, an upper bound on the sum of the complexity estimation of the $(d)$-1 step for \textcolor{black}{all} $2 \le d \le D$ is given by
\begin{eqnarray*}
\label{eq:D1}
	\sum_{d=2}^{D} \max \{|\tilde{\mathscr{S}}_d^{(k)}|, |\mathscr{T}_d^{(k)}| \} ^\omega
	&\le& \left( \sum_{d=2}^{D} \max \{|\tilde{\mathscr{S}}_d^{(k)}|, |\mathscr{T}_d^{(k)}| \} \right)^\omega \le \left( |\mathscr{T}_{\leq D}^{(k)}| + \alpha \right)^\omega \nonumber \\
	&\le& (2\cdot|\mathscr{T}_{\le D}^{(k)}|) ^\omega
	= O \left( \tbinom{n-k+D}{D}^\omega \right),
\end{eqnarray*}
where we used the equality
\[
\sum_{d=2}^{D} \max \{|\tilde{\mathscr{S}}_d^{(k)}|, |\mathscr{T}_d^{(k)}| \} = \sum_{d=2}^D  \max\{|\tilde{\mathscr{S}}_d^{(k)}| -|\mathscr{T}_d^{(k)}|, 0\} + |\mathscr{T}_{\leq D}^{(k)}|.
\]
Therefore, we set $C_{(d)1}$ to be $\tbinom{n-k+D}{D}^\omega$.

%=================================
\paragraph*{Time Complexity of $(d)$-2.}
%=================================

In each $(d)$-2 step, the complexity of executing the same row operations as those in $(d)$-1 step
is estimated as that of multiplying \textcolor{black}{a square matrix over $\mathbb{F}_q$ of size $|\tilde{\mathscr{S}}_d^{(k)}|\times |\tilde{\mathscr{S}}_d^{(k)}|$} to the \textcolor{black}{polynomial} matrix \textcolor{black}{$\mathcal{PM}[\tilde{\mathscr{S}}_d^{(k)},\mathscr{T}_{(d-2);(d-1)}^{(k)}]$ from the left.
Note that $\mathcal{PM}[\tilde{\mathscr{S}}_d^{(k)},\mathscr{T}_{(d-2);(d-1)}^{(k)}]$} is a sparse matrix, \textcolor{black}{since $\mathcal{PM}[\mathscr{S}_d^{(k)},\mathscr{T}_{\le (d-1)}^{(k)}]$ does not change from the original structure at the beginning of the ($d$)-2 step by the same discussion as in Subsection~\ref{sub:detail}}, where each row of it has at most $n-k+1$ non-zero entries.
Thus, multiplying the two matrices are done \textcolor{black}{in $O((n-k)\cdot|\tilde{\mathscr{S}}_d^{(k)}|^2)$} additions and scalar multiplications
in $\mathbb{F}_q[x_1,\dots,x_k]$.
Since polynomials appearing in each addition or scalar multiplication have degree $\le 2$,
its cost is bounded by $O\left(\tbinom{k+2}{2}\right)$ with naive approach.
Considering above together, each $(d)$-2 step has complexity \textcolor{black}{$O\left( \tbinom{k+2}{2}\cdot(n-k)\cdot|\tilde{\mathscr{S}}_d^{(k)}|^2\right)$}, and
hence the total complexity of $(d)$-2 for all $2 \le d \le D$ is given by
%
% At the time of executing ($d$)-2, the matrix $\mathcal{PM}[\tilde{I}_d,T_{(d-2);(d-1)}]$ is sparse from the same discussion in Subsection~\ref{sub:detail}.
% Indeed, each row of $\mathcal{PM}[\tilde{I}_d,T_{(d-2);(d-1)}]$ has at most $n-k+1$ nonzero elements.
% Furthermore, the degree of every element is smaller than or equal to 2,
% and we here estimate by $O(\tbinom{k+2}{2})$ the complexities of addition of polynomials in $\mathbb{F}_q[x_1,\dots,x_k]$ with degree $\le 2$ and multiplication between such a polynomial and an element of $\mathbb{F}_q$ with a naive approach.
% By considering these facts, an upper bound on the sum of the complexity of the $(d)$-2 step over $\mathbb{F}_q$ for $2 \le d \le D$ is
%
\begin{eqnarray*}
\label{eq:D2}
	\sum_{d=2}^{D} \left( \tbinom{k+2}{2}\cdot \left(n-k \right)\cdot |\tilde{\mathscr{S}}_d^{(k)}|^2 \right) & \leq & \tbinom{k+2}{2}\cdot \left(n-k \right) \cdot\left( \sum_{d=2}^{D} |\tilde{\mathscr{S}}_d^{(k)}| \right)^2 \\
	& = & \tbinom{k+2}{2} \cdot \left(n-k \right)\cdot |\mathscr{T}_{\le D}^{(k)}|^2 \nonumber \\
	&=& O \left( k^2 \cdot \left(n-k \right)\cdot \tbinom{n-k+D}{D}^2 \right) ,
\end{eqnarray*}
and thus $C_{(d)2}$ is set to be $k^2 \cdot \left(n-k \right)\cdot \tbinom{n-k+D}{D}^2$.

%================================
\paragraph*{Time Complexity of $(d)$-3.}
%================================
To estimate the time complexity of $(d)$-3 \textcolor{black}{for all $d$} with $2 \le d \le D$, we use the following lemma:
\begin{lemma}
\label{lem:d3_deg}
	At the time of executing the $(d)$-3 step with $2 \le d \le D-1$, the degree of every element of \textcolor{black}{$\mathcal{PM}[\tilde{\mathscr{S}}_{(d+1);D}^{(k)},\mathscr{T}_d^{(k)}]$} is lower than or equal to $D-d$.
\end{lemma}
\begin{proof}
	By the induction, we prove that,
	at the time of \textcolor{black}{starting} the $(d)$-3 step, the degree of every element of \textcolor{black}{$\mathcal{PM}[\tilde{\mathscr{S}}_{(d+1);D}^{(k)},\mathscr{T}_{d}^{(k)}]$ and $\mathcal{PM}[\tilde{\mathscr{S}}_{(d+1);D}^{(k)},\mathscr{T}_{d-1}^{(k)}]$} is lower than or equal to $D-d$ and $D-d+1$, respectively.
	In the case \textcolor{black}{of} $d=D-1$, the above statement clearly holds.
	In the following, we show that, if the statement holds when $d=d'$ with $3 \le d' \le D-1$, then it also holds when $d=d'-1$.
	Before executing the step $(d')$-3, \textcolor{black}{it is clear that $\mathcal{PM}[\tilde{\mathscr{S}}_{(d'+1);D}^{(k)},\mathscr{T}_{d'-2}^{(k)}]$} is a zero matrix.
	Then, the $(d')$-3 step adds row vectors, which are obtained by multiplying rows corresponding to \textcolor{black}{$\tilde{\mathscr{S}}_{d'}^{(k)}$} by a polynomial with the degree $D-d'$, to rows corresponding to \textcolor{black}{$\tilde{\mathscr{S}}_{(d'+1);D}^{(k)}$}.
	Here, the degree of each entry of \textcolor{black}{$\mathcal{PM}[\tilde{\mathscr{S}}_{d'}^{(k)},\mathscr{T}_{d'-1}^{(k)}]$ and $\mathcal{PM}[\tilde{\mathscr{S}}_{d'}^{(k)},\mathscr{T}_{d'-2}^{(k)}]$} are at most 1 and 2, respectively.
	Hence, through $(d')$-3, the degree of each entry of \textcolor{black}{$\mathcal{PM}[\tilde{\mathscr{S}}_{(d'+1);D}^{(k)},\mathscr{T}_{d'-2}^{(k)}]$} becomes at most $D-d'+2$ and that of \textcolor{black}{$\mathcal{PM}[\tilde{\mathscr{S}}_{(d'+1);D}^{(k)},\mathscr{T}_{d'-1}^{(k)}]$} remains at most $D-d'+1$, 
	Therefore, the statement holds in the case where $d=d'-1$, as desired.
 \qed
\end{proof}
% \noindent
% A proof is given in that of \cite[Lemma 10]{FK22}, and for reasons of space, we omit to write it here.

%\noindent
Each $(d)$-3 step eliminates the corresponding columns using the leading coefficients of \textcolor{black}{$\mathcal{PM}[\tilde{\mathscr{S}}_{d}^{(k)},\mathscr{T}_{d}^{(k)}]$}.
\textcolor{black}{More concretely, for each $i$ with $1 \leq i \leq r_d$, we conduct row operations to eliminate the non-zero entries in the column to which the leading coefficient of the $i$-th row of $\mathcal{PM}[\tilde{\mathscr{S}}_{d}^{(k)},\mathscr{T}_{d}^{(k)}]$ (in reduced row echelon form) belong, where $r_d$ is the rank of $\mathcal{PM}[\tilde{\mathscr{S}}_{d}^{(k)},\mathscr{T}_{d}^{(k)}]$.
Such the non-zero entries to be eliminated are ones of $\mathcal{PM}[\tilde{\mathscr{S}}_{(d+1);D}^{(k)},\mathscr{T}_{d}^{(k)}]$, and we suppose from \eqref{eq:row_mac} that the number of them is at most $\alpha$ for each $i$.
In each elimination process, we multiply the $i$-th row of $\mathcal{PM}[\tilde{\mathscr{S}}_{d}^{(k)},\mathscr{T}_{(d-2);d}^{(k)}]$ by a non-zero polynomial in $\mathbb{F}_q[x_1,\ldots,x_k]$ of degree at most $D-d$ (this degree bound comes from Lemma~\ref{lem:d3_deg}), and then add the multiple to a row of $\mathcal{PM}[\tilde{\mathscr{S}}_{(d+1);D}^{(k)},\mathscr{T}_{(d-2);d}^{(k)}]$.
Since each entry of $\mathcal{PM}[\tilde{\mathscr{S}}_{d}^{(k)},\mathscr{T}_{(d-2);d}^{(k)}]$ is a polynomial in $\mathbb{F}_q[x_1,\ldots,x_k]$ of degree $\leq 2$ at this point, and since $\mathcal{PM}[\tilde{\mathscr{S}}_{d}^{(k)},\mathscr{T}_{(d-2);d}^{(k)}]$ has $|\mathscr{T}_{(d-2);d}^{(k)}| = O(|\mathscr{T}_{d}^{(k)}|)$ columns, each elimination process is done in $O\left(\tbinom{k+2}{2}\cdot\tbinom{k+D-d}{D-d} \cdot |\mathscr{T}_{d}^{(k)}|\right)$ with a naive approach.
The total number of these elimination processes is upper-bounded by $r_d \cdot \alpha$,}
we estimate the complexity of the $(d)$-3 step as
\begin{equation*}
	O\left( \tbinom{k+D-d}{D-d} \cdot \tbinom{k+2}{2} \cdot \alpha \cdot r_d \cdot |\mathscr{T}_d^{(k)}| \right) \le O\left( \tbinom{k+D-d}{D-d} \cdot \tbinom{k+2}{2} \cdot \alpha \cdot \tbinom{n-k+d-1}{d} ^2 \right).
\end{equation*}
Note that the $(D)$-3 step can be omitted since \textcolor{black}{$\mathcal{PM}[\tilde{\mathscr{S}}_{\le (D-1)}^{(k)},\mathscr{T}_D^{(k)}]$} is a zero matrix.
Consequently, the sum of the \textcolor{black}{complexities} of the $(d)$-3 step for \textcolor{black}{all $d$ with} $2 \le d \le D-1$ is estimated by
\begin{eqnarray}
\label{eq:D3}
	&&\sum_{d=2}^{D-1} \left( \tbinom{k+D-d}{D-d} \cdot \tbinom{k+2}{2} \cdot \alpha \cdot \tbinom{n-k+d-1}{d} ^2 \right) \nonumber \\
	&&\le \tbinom{k+2}{2} \cdot \alpha \cdot \left( \sum_{d=2}^{D-1} \tbinom{n-k+d-1}{d} \right) \cdot \left( \sum_{d=2}^{D-1} \tbinom{k+D-d}{k} \cdot \tbinom{n-k+d-1}{n-k-1} \right).
\end{eqnarray}
Putting $d' = k+D-d$, one has
\begin{eqnarray*}
    &&\sum_{d=2}^{D-1} \tbinom{k+D-d}{k} \cdot \tbinom{n-k+d-1}{n-k-1}   =  \sum_{d'=k+1}^{k+D-2} \tbinom{d'}{k}\tbinom{(n+D-1)-d'}{(n-1)-k}\\
    & \leq & \sum_{d'=0}^{n+D-1} \tbinom{d'}{k}\tbinom{(n+D-1)-d'}{(n-1)-k} = \tbinom{(n+D-1)+1}{(n-1)+1} = \tbinom{n+D}{D}
\end{eqnarray*}
from a formula similar to Vandermonde's identity.
Therefore, the right hand side of \eqref{eq:D3} is upper-bounded by
\[
O\left( k^2 \cdot \alpha \cdot \tbinom{n-k+D}{D} \cdot \tbinom{n+D}{D} \right),
\]
and thus we set $C_{(d)3}$ to be $k^2 \cdot \alpha \cdot \tbinom{n-k+D}{D} \cdot \tbinom{n+D}{D}$.

%====================================
\paragraph*{Time Complexity of \textbf{Fix}.}
%====================================
The size of the resulting matrix of \textbf{Linearize(1)} is approximately $\alpha \times \alpha$ due to the discussion in Subsection~\ref{sub:alpha}, and
the degree of every element in the matrix is lower than or equal to $D$ from Lemma~\ref{lem:d3_deg}.
Therefore, the time complexity of \textbf{Fix} is estimated as that of substituting $k$ values to $x_1, \dots, x_k$ in $\alpha^2$ polynomials with degree $D$ in $\mathbb{F}_q[x_1,\dots,x_k]$.
When we use a naive approach, the complexity of evaluation of a polynomial with degree $d$ in \textcolor{black}{$k$} variables is estimated by \textcolor{black}{$\tbinom{k+d}{d}$}.
Therefore, $C_{\mathsf{fix}}$ is given by
\begin{equation}
\label{eq:subs}
	C_{\mathsf{fix}} = q^k \cdot \alpha^2 \cdot \tbinom{k+D}{D},
\end{equation}
since the \textbf{Fix} step is iterated for any values of $x_1, \dots, x_k$.

%===========================================
\paragraph*{Time Complexity of \textbf{Linearize(2)}.}
%===========================================

The \textbf{Linearize(2)} step performs Gaussian elimination on an $\alpha \times \alpha$ matrix over $\mathbb{F}_{q}$, and thus we estimate $C_{\mathsf{li2}}$ by
\begin{equation}
\label{eq:linear2}
	C_{\mathsf{li2}} = q^k \cdot \alpha^\omega,
\end{equation}
considering $q^k$ times iterations.

%=============================================
\subsubsection*{Rough Estimations of Time Complexity}
%=============================================
Here, we present a more compact formula for the time complexity of PXL.
Comparing the estimations $C_{(d)2}$ and $C_{(d)3}$, we can easily confirm that the value of $C_{(d)3}$ is larger than that of $C_{(d)2}$.
Furthermore, comparing the estimations $C_{(d)1}$ and $C_{(d)3}$, we experimentally confirmed that, for the case where $10 \le n\le 100$, $m=n,\,1.5n,\,2n$, and $k$ is the value minimizing the sum of the above five estimations, the value of $C_{(d)3}$ is always much larger than that of $C_{(d)1}$
(e.g., $C_{(d)1}$ and $C_{(d)3}$ \textcolor{black}{in} the case \textcolor{black}{where} $n=m=100$ with $q=2^8$ is approximately $2^{210}$ and $2^{259}$, respectively).
These facts indicate that the complexity of the \textbf{Linearize(1)} step is dominated by $C_{(d)3}$ for \textcolor{black}{practical} cases, and it is estimated as follows:
\begin{equation}
\label{eq:linear1}
	O\left( k^2 \cdot \alpha \cdot \tbinom{n-k+D}{D} \cdot \tbinom{n+D}{D} \right).
\end{equation}
By using this \textcolor{black}{estimation on $C_{(d)3}$}, the time complexity of PXL is roughly estimated by $C_{(d)3}+C_{\mathsf{fix}}+C_{\mathsf{li2}}$, say
\begin{equation}
\label{eq:pro_odd}
	O\left( k^2 \cdot \alpha \cdot \tbinom{n-k+D}{D} \cdot \tbinom{n+D}{D} + q^k \cdot \left( \alpha^2 \cdot \tbinom{k+D}{D} + \alpha^\omega \right) \right).
\end{equation}

\begingroup
\renewcommand{\arraystretch}{1.2}
\begin{table}[t]
	\centering
	\caption{The number of field operations approximated by power of 2 between PXL~\eqref{eq:pro_odd}, h-XL~\eqref{eq:hyb_XL}, h-WXL~\eqref{eq:hyb_WXL}, and Crossbred~\cite{estimator}, the optimal number $k$ of guessed variables of PXL, \textcolor{black}{the value of $D = D_1^{(n-k)}$ estimated in \eqref{eq:Dreg_k}}, and the estimated size $\alpha$ of the resulting matrix of \textbf{Linearize(1)} on the MQ system with $n=m=20$, $40$, $60$, and $80$ over ${\mathbb F}_{2^8}$ (above) and over ${\mathbb F}_{31}$ (below).}
	\label{tab:comp}

	\vspace{5pt}

	\begin{tabular}{wc{0.5cm}|wc{1.7cm}|wc{1.1cm}wc{1.1cm}wc{1.1cm}wc{1.1cm}wc{1.1cm}wc{1.1cm}wc{1.1cm}wc{1.1cm}}
		\hline
		\multirow{9}{*}{$\mathbb{F}_{2^8}$} &$n=m$& \multicolumn{2}{c}{$20$} & \multicolumn{2}{c}{$40$} & \multicolumn{2}{c}{$60$} & \multicolumn{2}{c}{$80$} \\
            &$\omega$ & 2.37 & 2.81 & 2.37 & 2.81 & 2.37 & 2.81 & 2.37 & 2.81  \\
\cline{2-10}
		&h-XL & $2^{75}$ & $2^{85}$ & $2^{134}$ & $2^{153}$ & $2^{194}$ & $2^{221}$ & $2^{252}$ & $2^{287}$ \\
		&h-WXL & $2^{75}$ & $2^{75}$ & $2^{129}$ & $2^{129}$ & $2^{182}$ & $2^{182}$ & $2^{234}$  & $2^{234}$ \\
            &Crossbred & $2^{65}$ & $2^{74}$ & $2^{123}$ & $2^{137}$ & $2^{180}$ & $2^{201}$ & $2^{237}$  & $2^{265}$ \\
		&\textbf{PXL} & $\mathbf{2^{62}}$ & $\mathbf{2^{64}}$ & $\mathbf{2^{117}}$ & $\mathbf{2^{121}}$ & $\mathbf{2^{169}}$ & $\mathbf{2^{178}}$ & $\mathbf{2^{220}}$  & $\mathbf{2^{233}}$ \\
		\cline{2-10}
		&$k$& 3 & 3 & 6 & 5 & 8 & 7 & 10  & 8 \\
		&$D$& 9 & 9 & 14 & 15 & 19 & 20 & 24  & 27 \\
		&$\alpha$& ${2^{14}}$ & $2^{14}$ & ${2^{27}}$ & $2^{29}$ & ${2^{42}}$ & $2^{44}$ & ${2^{56}}$  & $2^{60}$ \\
		\hline
	\end{tabular}

	\vspace{5pt}	

	\begin{tabular}{wc{0.5cm}|wc{1.7cm}|wc{1.1cm}wc{1.1cm}wc{1.1cm}wc{1.1cm}wc{1.1cm}wc{1.1cm}wc{1.1cm}wc{1.1cm}}
		\hline
		\multirow{9}{*}{$\mathbb{F}_{31}$} &$n=m$ & \multicolumn{2}{c}{$20$} & \multicolumn{2}{c}{$40$} & \multicolumn{2}{c}{$60$} & \multicolumn{2}{c}{$80$} \\
            &$\omega$ & 2.37 & 2.81 & 2.37 & 2.81 & 2.37 & 2.81 & 2.37 & 2.81  \\
				\cline{2-10}
		&h-XL & $2^{66}$ & $2^{73}$ & $2^{119}$ & $2^{131}$ & $2^{170}$ & $2^{191}$ & $2^{221}$  & $2^{246}$ \\
		&h-WXL & $2^{65}$ & $2^{65}$ & $2^{116}$ & $2^{116}$ & $2^{162}$ & $2^{162}$ & $2^{208}$  & $2^{208}$ \\
            &Crossbred & $2^{57}$ & $2^{62}$ & $2^{109}$ & $2^{117}$ & $2^{158}$ & $2^{170}$ & $2^{208}$ & $2^{224}$ \\
		&\textbf{PXL} & $\mathbf{2^{57}}$ & $\mathbf{2^{57}}$ & $\mathbf{2^{105}}$ & $\mathbf{2^{107}}$ & $\mathbf{2^{152}}$ & $\mathbf{2^{158}}$ & $\mathbf{2^{197}}$ & $\mathbf{2^{208}}$ \\
		\cline{2-10}
		&$k$& 5 & 5 & 8 & 8 & 11 & 10 & 13 & 12 \\
		&$D$& 7 & 7 & 12 & 12 & 16 & 17 & 21 & 22 \\
		&$\alpha$& ${2^{11}}$ & $2^{11}$ & ${2^{24}}$ & $2^{24}$ & ${2^{37}}$ & $2^{38}$ & ${2^{51}}$ & $2^{53}$ \\
		\hline
	\end{tabular}

\end{table}
\endgroup

%======================================
\subsection{Comparison}\label{sub:time_comp}
%======================================

We compare the complexity of our PXL with those of h-XL, h-WXL, and Crossbred with our motivation towards contribution of PXL to evaluating the security of MPKCs.
Following the security estimation of \cite{UOV}, we choose h-WXL among the XL family as a target for comparison.
We also \textcolor{black}{adopt} the complexity of h-XL on which h-WXL is originally based (in fact, h-XL is the most basic method in the framework of the hybrid approaches with XL) and \textcolor{black}{that of} Crossbred recognized as the theoretical most efficient algorithm for some parameter sets in~\cite{BMS22}.
%Since the mo
%
%Following 
%The reason why we choose h-XL and h-WXL as targets for comparison is the following:
%Among the XL family, the complexity estimation of h-WXL is currently used to estimate the security level of MPKCs~\cite{Rainbow3}, and h-XL is constructed by applying the hybrid approach to the plain XL.
%
Recall that the complexities of h-XL, h-WXL, Crossbred, and PXL are estimated by \eqref{eq:hyb_XL}, \eqref{eq:hyb_WXL}, \cite{estimator} and \eqref{eq:pro_odd}, respectively, {where the estimation~\eqref{eq:pro_odd} for our PXL is obtained by supposing practical \textcolor{black}{Assumptions \ref{assump:semireg_eval} and \ref{assump:q_semireg}, and Heuristic \ref{heur:choose}}.}
Note that, for fixed $n$, $m$, and $q$, each of the \textcolor{black}{four} approaches chooses the number $k$ of guessed variables (and $D$ and $d$ for Crossbred) so that its complexity estimation becomes the smallest value, and thus the value of $k$ depends on each approach.
%Furthermore, we here use $\omega = 2.37$ following~\cite{Gal14}.
Furthermore, we here take \textcolor{black}{the exponent of matrix multiplication $\omega$} as $2.37$~\cite{Gal14} and $2.81$~\cite{Str69}.
As we will see below, PXL is theoretically more efficient than other algorithms in the case of $n=m$ (this is the case \textcolor{black}{where} hybrid approaches for the MQ problem \textcolor{black}{work} most efficiently).

Table~\ref{tab:comp} compares the bit complexities of PXL, h-XL, h-WXL, and Crossbred on the MQ system of $m$ equations in $n$ variable with $n=m$ over ${\mathbb F}_{2^8}$ and ${\mathbb F}_{31}$.
These orders of the finite fields are chosen following the MQ challenge~\cite{MQcha}, and in particular, $q=2^8=256$ is also suggested as a parameter of \cite{UOV}. 
Note also that we do not choose $q=2$ since exhaustive searches are known to be effective in this case. 
Specifically, Table~\ref{tab:comp} shows the bit complexities of the four approaches, the optimal $k$ of PXL minimizing the value of \eqref{eq:pro_odd}, \textcolor{black}{the value of $D = D_1^{(n-k)}$ estimated in \eqref{eq:Dreg_k}}, and the estimated size $\alpha$ of the resulting matrix of \textbf{Linearize(1)} obtained from \eqref{eq:alpha}
for the case where $n=m$ \textcolor{black}{with $n \in \{ 20,40, 60, 80\}$}.
%(We have also compared the complexities over other finite fields, which are omitted here for reasons of space.)
For example, \textcolor{black}{when} $q=2^8$, $n=m=80$, and $\omega=2.37$, the complexities of h-XL, h-WXL, Crossbred, and PXL are approximately estimated as $2^{252}$, $2^{234}$, $2^{237}$, and $2^{220}$, respectively.
As a result, we \textcolor{black}{expect} that PXL has the less complexity than those of other algorithms especially in the case of $\omega=2.37$; we \textcolor{black}{also} expect that \textcolor{black}{similar} results will be obtained in other finite fields from the form of the complexity estimation~\eqref{eq:pro_odd}.

On the other hand, we confirmed that PXL is not efficient in highly overdetermined cases.
This is because, in such overdetermined cases, $k$ is set to be a very small value for efficiency.
%We confirmed that such behavior of PXL can be seen over other finite fields.

\begin{remark}[Space Complexity]
The memory space consumed by PXL is upper-bounded by $O \left( \tbinom{k+D}{D} \cdot \tbinom{n-k+D}{D}^2 \right)$, since the degree of every element of the Macaulay matrix and its transformed matrices in \textbf{Linearize(1)} is at most $D$ through an execution of PXL from Lemma~\ref{lem:d3_deg}.
This estimation cannot be directly compared with other algorithms, since the values of the following two parameters depend on one's choice of an algorithm: \textcolor{black}{The degree bound $D$ (for the success of the algorithm)} and the number $k$ of fixed values.

On the other hand, focusing on the sparsity/density of matrices, we predict that PXL is not efficient compared with h-WXL in terms of the space complexity for the following reason:
Through the elimination process of Macaulay matrices, WXL can deal with a Macaulay matrix as a sparse matrix due to Wiedemann's algorithm, whereas PXL \textcolor{black}{maintains} some dense submatrices.
Considering this together with the time complexities for practical parameters, we conclude that the relationship between PXL and h-WXL would be a trade-off between time and memory.
\end{remark}

%=================
\section{Experimental Results}
%=================

\begin{figure}[t]
	\centering
	\includegraphics[width=10.0cm]{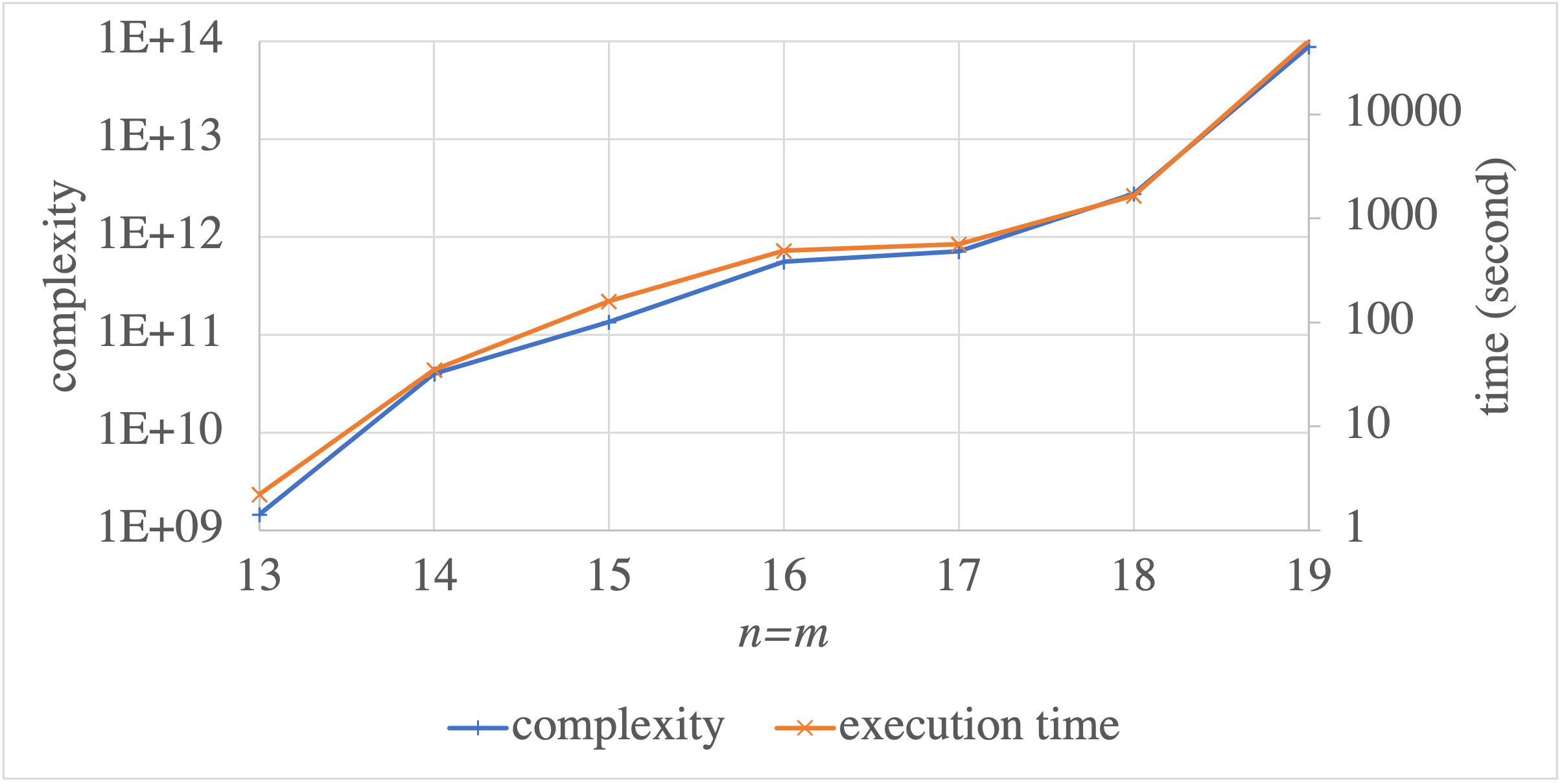}
	
	\caption{Comparison between the estimation of complexity by~\eqref{eq:linear1} and the execution time of the \textbf{Linearize(1)} step on an MQ system with $n=m$ over ${\mathbb F}_{2^4}$.}
	\label{fig:linear1}

\end{figure}

We implemented the proposed algorithm PXL in the Magma computer algebra system (V2.26-10)~\cite{Magma}, in order to examine that it behaves as our complexity estimation provided in Section \ref{sec:complexity}.
(As it will be described below, note that our current implementation is not optimized one, see also Remark \ref{rem:exp}.)
%(and thus such a comparison is not fair).
We also confirmed in our experiments that PXL outputs a solution correctly at \textcolor{black}{$D = D_1^{(n-k)}$ as estimated in \eqref{eq:Dreg_k}}.
%Concretely, the \textbf{Multiply} and \textbf{Linearize(1)} steps are implemented over Magma, and the other steps are implemented over C.
%Among the six steps of PXL described in Algorithm~\ref{alg:PXL}, the last four steps from \textbf{Fix} to \textbf{Repeat} can be constructed using existing optimized implementations.

%In this appendix, we mainly discuss the implementation of the \textbf{Multiply} and \textbf{Linearize(1)} steps
%(in particular, the last four steps from \textbf{Fix} to \textbf{Repeat} can be more optimized).		
First, we confirmed that the \textbf{Linearize(1)} step behaves as in~\eqref{eq:linear1}.
%, see~\cite[Appendix A]{FK22} for details.
The reason why we focus on the behavior of the \textbf{Linearize(1)} step is the following:
In the estimation \eqref{eq:pro_odd} of the total time complexity, only $C_{(d)3}$ is specific to our estimation in theory, while the later parts $C_{\mathsf{fix}}$ and $C_{\mathsf{li2}}$ for the \textbf{Fix} and \textbf{Linearize(2)} steps just come from \textcolor{black}{known complexity estimations}.
%First, we compare the execution time of the \textbf{Multiply} and \textbf{Linearize(1)} steps of our implementation over Magma and the number of manipulations estimated by~\eqref{eq:linear1}.
Figure~\ref{fig:linear1} compares the execution time of the \textbf{Linearize(1)} step and the bit complexity \eqref{eq:linear1} on the system with $n=m$ from $n=13$ to $n=19$ over ${\mathbb F}_{2^4}$,
and the number $k$ of fixed variables is chosen so as to minimize the value of~\eqref{eq:pro_odd}.
As a result, Figure~\ref{fig:linear1} shows that the execution time and our estimation~\eqref{eq:linear1} have almost the same behavior,
which indicates that the estimation~\eqref{eq:linear1} would be reliable. 

On the other hand, our current Magma implementation of the \textbf{Fix} and \textbf{Linearize(2)} steps does not show the similar behavior as our complexity estimation, due to the use of unoptimized implementation.
For example, in the case of $n=m=16$ with $k=5$, \textbf{Linearize(1)}, \textbf{Fix}, and \textbf{Linearize(2)} took 10~min., 40~hr., and 30~min., respectively,
whereas the estimated numbers of field operations of these three steps from \eqref{eq:linear1}, \eqref{eq:subs}, and \eqref{eq:linear2} are $2^{39}$, $2^{44}$, and $2^{39}$, respectively.
We observe that this inefficiency of the latter two steps (in particular \textbf{Fix} with a lot of for-loops) is due to the use of Magma's interpreter language.
Using compiler languages such as C instead could be a solution to resolve this problem, but we must newly implement the arithmetic of matrices and polynomials efficiently, which is not the topic of this paper.
We leave such an efficient implementation with compiler languages to future work.

\begin{remark}\label{rem:exp}
We remark that here we do not compare the execution time of our PXL with that of any other variant of XL, since the practical behavior deeply depends on how one implements the arithmetic of matrices (and polynomials) efficiently, which is not the topic of this paper.
For a fair comparison, providing optimized implementations of several variants including PXL is required, and it is a very important task for practical cryptanalysis.
\end{remark}

%=================
\section{Conclusion}
%=================
We presented a new variant of XL, which is a major approach for solving the MQ problem.
Our proposed polynomial XL (PXL) eliminates the linearized monomials in polynomial rings to solve the system efficiently, and we estimated its complexities.
Given an MQ system of $m$ equations in $n$ variables, the proposed algorithm first regards each polynomial in $n$ variable as \textcolor{black}{one} in $n-k$ variables $x_{k+1},\dots, x_n$, whose coefficients belong to the polynomial ring ${\mathbb F}_q[x_1,\dots, x_k]$.
We then generate a Macaulay matrix over ${\mathbb F}_q[x_1,\dots, x_k]$, and partly perform the row reduction (Gaussian elimination).
Finally, random values are substituted for the $k$ variables, and the remaining part of the (partly-reduced) Macaulay matrix is transformed into the reduced row echelon form.
Partly reducing the (polynomial) Macaulay matrix is done mainly on submatrices over $\mathbb{F}_q$ (not over ${\mathbb F}_q[x_1,\dots, x_k]$) with arithmetic of polynomials in ${\mathbb F}_q[x_1,\dots, x_k]$ of bounded degree, and under some \textcolor{black}{practical assumption and heuristic} \textcolor{black}{(Assumption \ref{assump:q_semireg} and Heuristic \ref{heur:choose})}, the remaining part is expected to have size much smaller than the original one.
This construction {can reduce} the amount of field operations for each guessed value, compared to h-XL.
Supposing the above assumption and heuristic and additional but still practical \textcolor{black}{one (Assumption \ref{assump:semireg_eval}), which assumes the affine semi-regularity of polynomial sequences}, \textcolor{black}{we gave} an asymptotic estimation of the time complexity \textcolor{black}{of} PXL, which implies that PXL could solve the system faster in theory for the case of $n \approx m$ than h-XL, h-WXL, and Crossbred.
On the other hand, PXL might be less efficient than h-WXL with respect to the space complexity.
%  for the following two reasons.
% : Macaulay matrices become denser through PXL and the resulting matrix of \textbf{Linearize(1)} has elements with degree higher than or equal to 2.

This paper discusses only the quadratic case, but, as in the plain XL, the proposed algorithm can be also generalized to higher degree cases.
Therefore, one considerable future work is to analyze the complexity of PXL on such higher degree systems.
Furthermore, for a comparison of the practical time-efficiencies of our PXL and other XL variants, it is important to implement PXL (and the other variants) efficiently.
%is also an important problem.
In our experiments, we implemented PXL over Magma, but this can be more optimized by using an alternative (compiler) programming language, e.g., C.
Note that there will be a drawback that the construction of our PXL over the polynomial ring prohibits the use of existing linear algebra libraries, which are often heavily optimized.
Therefore, to provide such an optimized code for PXL will be a challenging task.
Finally, we leave the analysis of the effect of PXL on \textcolor{black}{the security of} various multivariate signature schemes to a future work.

%%%%%%%%%%%%%%%%%%%%%%%%%%%%%%%%%%%%%%%%%%%%%%%%%%%%%%%%%%%%%%%%%%%%%%%%%%%%%%%%%%%%%%%%%%%%%%%%%%%%%% 

\subsection*{Acknowledgements}
The authors thank the anonymous referees for helpful comments and suggestions.
The authors also thank Tsuyoshi Takagi and Kazuhiro Yokoyama for helpful comments and suggestions.
The authors are grateful to Kosuke Sakata for his advice on the implementation of our proposed algorithm. 

This work was supported by
JST CREST Grant Number JPMJCR2113, Japan,
JSPS KAKENHI Grant Number JP22KJ0554, Japan, and
JSPS Grant-in-Aid for Young Scientists 20K14301 and 23K12949, Japan.

%%%%%%%%%%%%%%%%%%%%%%%%%%%%%%%%%%%%%%%%%%%%%%%%%%%%%%%%%%%%%%%%%%%%%%%%%%%%%%%%%%%%%%%%%%%%%%%%%%%%%% 

\bibliographystyle{plain}
\bibliography{sample}

\appendix

\section{Semi-regular sequences}\label{app:semireg}

We here review the notion of {\it semi-regular} sequence, which is introduced first by Bardet et al.\ (e.g., \cite{B04}, \cite{BFS}, \cite{BFSY}).
%motivated to analyze the complexity of computing Gr\"{o}bner bases.
Semi-regular sequences are formulated also by Diem~\cite{Diem2} in terms of commutative and homological algebra.
\textcolor{black}{See also \cite[Section~2]{KY} for a survey.}

We use the following notation:
Let $R = K[x_1,\ldots,x_n]$ be the polynomial ring of $n$ variables $x_1,\ldots, x_n$ over a field $K$.
For a finitely generated graded $R$-module $M = \bigoplus_{d \in \mathbb{Z}}M_d$ \textcolor{black}{(namely $M_d$ is the degree-$d$ homogeneous component)}, we denote by ${\rm HF}_M$ its Hilbert function, namely ${\rm HF}_{M}(d) = \mathrm{dim}_K M_d$ for each integer $d$, and denote by ${\rm HS}_M$ the Hilbert series of $M$, say ${\rm HS}_{M}(z) = \sum_{d=0}^{\infty} {\rm HF}_{M}(d) z^d \in \mathbb{Z} \llbracket z \rrbracket$.
For a sequence $(f_1,\ldots, f_m)$ of {\it homogeneous} polynomials in $R$ of positive degrees, let $K_{\bullet}(f_1,\ldots,f_m)$ denote the Koszul complex on the sequence \textcolor{black}{(see e.g., \cite[Section 7.6]{Singular} for its definition)}, and let $H_i(K_{\bullet}(f_1,\ldots,f_m))$ be its $i$-th homology group.
In particular, the first homology group is a finitely generated graded $R$-module given by
\begin{equation}\label{eq:H1}
    H_1(K_{\bullet}(f_1,\ldots,f_m)) = \mathrm{syz}(f_1,\ldots , f_m)/ \mathrm{tsyz}(f_1,\ldots , f_m),
\end{equation}
\textcolor{black}{the sum of whose homogeneous components of degree less than or equal to $d$ is denoted by $H_1(K_{\bullet}(f_1,\ldots,f_m))_{\leq d}$ for each $d \in \mathbb{Z}$.}
Here, $\mathrm{syz}(f_1,\ldots,f_m)$ denotes the module of syzigies on $(f_1,\ldots, f_m)$, say
\[
\mathrm{syz}(f_1,\ldots,f_m) = \left\{ (h_1,\ldots , h_m) \in \bigoplus_{j=1}^m R(-d_j) \mathbf{e}_j\right\},
\]
where each $R(-d_j)$ is the shifted graded ring given by $R(-d_j)_d = R_{d-d_j}$ for $d \in \mathbb{Z}$, and where each $\mathbf{e}_j$ denotes a standard basis element.
On the other hand, \textcolor{black}{$\mathrm{tsyz}(f_1,\ldots,f_m)$} is defined as \textcolor{black}{an} $R$-submodule of $\mathrm{syz}(f_1,\ldots,f_m)$ given by
\begin{equation*}
    \mathrm{tsyz}(f_1,\ldots ,f_m) := \langle \mathbf{t}_{i,j}:= f_i \mathbf{e}_j - f_j \mathbf{e}_{i} : 1 \leq i < j \leq m \rangle_R,
\end{equation*}
which is called the module of trivial syzigies on $(f_1,\ldots,f_m)$.

% The terminology ``cryptographic'' was named by Bigdeli et al.\ in their recent work~\cite{BNDGMT21}, in order to distinguish such a sequence from a semi-regular one defined by Pardue (see Definition \ref{def:semireg}).

We first recall the definiton of $d$-regular sequences:

\begin{definition}[{\cite[Definition 3]{BFS}}, {\cite[Definition 1]{Diem2}}]\label{def:semiregB}
Let $f_1, \ldots , f_m \in R$ be homogeneous polynomials of positive degrees $d_1, \ldots , d_m$ respectively, and put $I = \langle f_1, \ldots , f_m \rangle_R$.
%The notations $I^{(i)}$ and $A^{(i)}$ are the same as in Subsection \ref{subsec:HilSemi}.
For each integer $d$ with $d \geq \mathrm{max}\{d_i : 1 \leq i \leq m \}$, we say that a sequence $(f_1, \ldots , f_m)$ is {\it $d$-regular} if it satisfies the following condition:
\begin{itemize}
    \item For each $i$ with $1 \leq i \leq m$, if a homogeneous polynomial $g \in R$ satisfies $g f_i  \in \langle f_1, \ldots , f_{i-1} \rangle_R$ and $\mathrm{deg}(g f_i) < d$, then we have $g \in \langle f_1, \ldots , f_{i-1} \rangle_R$.
    % In other word, the multiplication map $A^{(i-1)}_{t-d_i} \longrightarrow A^{(i-1)}_{t} \ ; \ g \mapsto g f_i$ is injective for every $t$ with $d_i \leq t < d$.
\end{itemize}
\end{definition}

The (truncated) Hilbert series of $d$-regular sequences was determined by Diem~\cite{Diem2}, as in the following proposition:

\begin{theorem}[cf.\ {\cite[Theorem 1]{Diem2}}]\label{lem:Diem2}
We use the same notation as in Definition \ref{def:semiregB}.
Then, the following are equivalent for each $d$ with $d \geq \mathrm{max}\{d_i : 1 \leq i \leq m \}$:
\begin{enumerate}
    \item The sequence $( f_1, \ldots , f_m )$ of homogeneous polynomials is $d$-regular.
    % Namely, for each $(i,t)$ with $1 \leq i \leq m$ and $d_i \leq t < d$, the equality $\dim_K A^{(i)}_{t} = \dim_K A^{(i-1)}_{t} - \dim_K A^{(i-1)}_{t-d_i}$ holds.
    % % \item For each $i$ with $1 \leq i \leq m$, we have
    % % \begin{equation}
    % %   {\rm HS}_{A^{(i)}}(z) \equiv {\rm HS}_{A^{(i-1)}}(z) (1-z^{d_i}) \pmod{z^d}.
    % % \end{equation}
    \item We have
    \begin{equation}\label{eq:dregHil}
    {\rm HS}_{R/\langle f_1,\ldots,f_m \rangle}(z) \equiv \frac{\prod_{j=1}^{m}(1-z^{d_j})}{(1- z)^n} \pmod{z^d}.
    \end{equation}
    \item $H_1 (K_{\bullet}(f_1, \ldots , f_m))_{\leq d-1} = 0$.
\end{enumerate}
\end{theorem}

% \begin{proposition}[{\cite[Proposition 2 (a)]{Diem2}}]\label{prop:Diem2}
% With the same notation as in Definition \ref{def:semiregB}, let $D$ and $i$ be natural numbers.
% Assume that {\color{black} $H_i (K(f_1,\ldots , f_m))_{\leq D} = 0$}.
% Then, for each $j$ with $1 \leq j < m$, we have 
% {\color{black} $H_i(K(f_1, \ldots , f_j))_{\leq D} = 0$}.
% \end{proposition}

Recall that a finitely generated graded $R$-module $M$ is said to be {\it Artinian} if there exists a sufficiently large $D \in \mathbb{Z}$ such that $M_d = 0$ for all $d \geq D$.

\begin{definition}[{\cite[Definition 4]{BFS}}, {\cite[{Definition 4}]{BFSY}}]\label{def:dreg}
For a homogeneous ideal $I$ of $R$, we define its {\it degree of regularity} $d_{\rm reg}(I)$ as follows:
If the finitely generated graded $R$-module $R/I$ is Artinian, we set $d_{\rm reg} (I) := \mathrm{min} \{ d : R_d = I_d \}$ with $I_d = I \cap R_d$, and otherwise we set $d_{\rm reg}(I) := \infty$.
\textcolor{black}{We also denote $d_{\rm reg}(I)$ by $d_{\rm reg}(F)$ for a subset \textcolor{black}{or a sequence} $F$ of homogeneous elements in $R$ generating \textcolor{black}{the homogeneous ideal} $I$.}
\end{definition}

% As for an upper-bound on the degree of regularity, we refer to \cite[Theorem 21]{GG23-2}.

% \begin{remark}
%     In Definition \ref{def:dreg}, since $R/I$ is Noetherian, it is Artinian if and only if it is of finite length.
%     In this case, the degree of regularity $d_{\rm reg}(I)$ is equal to the {\it Castelnuovo-Mumford regularity} $\mathrm{reg}(I)$ of $I$ (see e.g., \cite[\S20.5]{Eisen} for the definition), whence $d_{\rm reg}(I) = \mathrm{reg}(I) = \mathrm{reg}(R/I) + 1$.
% \end{remark}

\begin{definition}[{\cite[Definition 5]{BFS}}, {\cite[Definition 5]{BFSY}}; see also {\cite[\S 2]{Diem2}}]\label{def:csemireg}
A sequence $(f_1, \ldots , f_m) \in R^m$ of homogeneous polynomials of positive degrees is said to be {\it semi-regular} if it is $d_{\rm reg}(I)$-regular, where we set $I = \langle f_1, \ldots , f_m \rangle_R$.
\end{definition}

The semi-regularity is characterized by equivalent conditions \textcolor{black}{in} the following proposition:

\begin{proposition}[{\cite[Proposition 1 (d)]{Diem2}}; see also {\cite[Proposition 6]{BFSY}}]\label{prop:Diem}
With the same notation as in Definition \ref{def:semiregB}, we put $D=d_{\rm reg}(I)$.
Then, the following are equivalent:
\begin{enumerate}
    \item The sequence $( f_1, \ldots , f_m )$ of homogeneous polynomials is semi-regular.
    \item We have
    \begin{equation}\label{eq:semiregHil2}
    {\rm HS}_{R/I}(z) = \left[ \frac{\prod_{j=1}^{m}(1-z^{d_j})}{(1- z)^n} \right],
    \end{equation}
    where $[\cdot]$ means truncating a formal power series over $\mathbb{Z}$ after the last consecutive positive coefficient.
    \item $H_1 (K_{\bullet}(f_1, \ldots , f_m))_{\leq D-1} = 0$.
\end{enumerate}
\end{proposition}

Note that, by Definition \ref{def:csemireg}, if $(f_1,\ldots,f_m)$ is semi-regular, then the degree of regularity $d_{\rm reg}(I)$ coincides with $\deg({\rm HS}_{R/I})+1$, where we set $I=\langle f_1,\ldots,f_m\rangle_R$. 

% In 1985, Fr\"{o}berg had already conjectured in \cite{Froberg} that, when $K$ is an infinite field, a generic sequence of homogeneous polynomials $f_1,\ldots,f_m \in R$ of degrees $d_1,\ldots , d_m$ generates an ideal $I$ with the Hilbert–Poincar\'{e} series of the form \eqref{eq:semiregHil2}, namely $(f_1,\ldots , f_m)$ is cryptographic semi-regular.
% It can be proved (cf.\ \cite{Pardue}) that Fr\"{o}berg's conjecture is equivalent to Pardue's one~\cite[Conjecture B]{Pardue}.
% We also note that Moreno-Soc\'{i}as conjecture~\cite{MS} is stronger than the above two conjectures, see \cite[Theorem 2]{Pardue} for a proof.

% It follows from the fourth condition of Proposition \ref{prop:semireg} together with the second condition of Proposition \ref{prop:Diem} that the semi-regularity implies the cryptographic semi-regularity.
% Note that, when $m\leq n$, both `semi-regular' and `cryptographic semi-regular' are equivalent to `regular'.

Finally, we recall the definition of an affine semi-regular sequence:

\begin{definition}[{\cite[Definition 5]{BFSY}}]\label{def:affine_semireg}
    A sequence ${F}= (f_1,\ldots , f_m)\in R^m$ of not necessarily homogeneous polynomials of positive degrees is said to be semi-regular if the sequence ${F}^{\rm top} = (f_1^{\rm top},\ldots, f_m^{\rm top})$ is semi-regular .
    In this case, the sequence $F$ is said to be {\it affine semi-regular}.
\end{definition}

\end{document}